\documentclass[floatfix,amsmath,aps,prb,eqsecnum]{revtex4-2}
\usepackage{bm}
\usepackage{xcolor}
\usepackage{hyperref}
\usepackage{makeidx}
\usepackage{amsmath}
\usepackage{graphicx}
\usepackage{dcolumn}
\usepackage{amsfonts}
\usepackage{amssymb}
\usepackage[latin9]{inputenc}
\usepackage{color}
\usepackage{subcaption}
\usepackage{float}

\begin{document}

\title{Interplay between topology and interactions in superconducting chains}

\author{A. C. P. Lima$^{1}$, M. S. Figueira$^2$, Mucio A. Continentino$^{1}$}

\email[corresponding author:]{muciocontinentino@gmail.com }
\affiliation{$^{1}$Centro Brasileiro de Pesquisas F\'{\i}sicas, Rua Dr. Xavier Sigaud, 150, Urca 22290-180, Rio de Janeiro, RJ, Brazil}
\affiliation{$^{2}$Instituto de F\'{i}sica, Universidade Federal Fluminense, Av. Litor\^anea s/N, CEP: 24210-340, Niter\'oi, RJ, Brasil}

\begin{abstract}
Most studies of non-trivial topological systems are carried out in non-interacting models that admit an exact solution. This raises the question, to which extent the consideration of electronic correlations and disorder, present in real systems, modify these results. Exact solutions  of correlated electronic systems with non-trivial topological properties, although fundamental are scarce. Among the non-interacting soluble models, we single out the Kitaev p-wave superconducting chain. It plays a crucial role in clarifying the appearance of emergent quasi-particles, the Majorana modes, associated with non-trivial topological properties. Given the relevance of this model, it would be extremely useful if it could be extended to include correlations and still remain solvable. In this work we investigate a superconducting Kitaev chain that interacts through a Falicov-Kimball Hamiltonian with a background of localized electrons. For some relevant values of the parameters,  this model can be solved exactly by mapping into a non-interacting one. This allows for a detailed study of the interplay between electronic correlations and non- trivial topological behavior. Besides, the random occupation of the chain by the local moments brings new interesting effects associated with disorder.

\end{abstract}

\maketitle

\section{Introduction}

Systems with non-trivial topological properties have been the subject of intensive studies in recent years~\cite{alicea,ando,review,nature,superspin,kitaev}. They have expanded the landscape  of interesting materials in the periodic table, giving rise to new concepts and to the discovery of unsuspected physical properties. Theoretical study of models with non-trivial topological properties has in many cases anticipated unusual behavior, creating exciting challenges to experimentalists. A clear example is the Kitaev superconducting chain model with emergent new quasi-particles, the Majorana fermions, on its edges~\cite{kitaev}. The unusual properties of these modes, with possible applications in quantum computers~\cite{Nayak1,Nayak2,Refael,Chou,Pan,Protocol} is leading to enormous progress in experimental nano-physics~\cite{alicea,review,nature,superspin,para2,v8,v9,tewari2,para3,bento,Robinson,Amitava,new2,tewari,desordem5}.

Many of the models of topological superconductors and insulators are idealized, since they do not include electronic correlations or disorder, and consequently can be solved exactly. In one aspect this is satisfactory, since it allows to obtain a clear picture of the nature of the emerging quasi-particles and topological non-trivial  properties. On the other hand, it raises the question to which extent these properties are robust to electronic correlations, or disorder,  present in 
real materials~\cite{rachel,desordem3,he,desordem2,desordem1,desordem0}. The motivation of the present study is to explore this problem presenting a non-trivial topological interacting model with an exact solution for some values of its parameters.

The Falicov-Kimball (FK) model plays an important role in the study of strongly correlated systems, as heavy fermions and mixed-valence material~\cite{brydon,novo,molecule}. The model consists of a conduction band of itinerant electrons interacting with a background of localized electrons. In this paper we obtain an exact solution for the one-dimensional spinless FK model with a $p$-wave pairing of the electrons in the conduction band.
The solution is possible since we  can map the many-body problem in a non-interacting system with a site dependent chemical potential~\cite{gotta1,derrida,gotta2,gotta2}. 
Our work allows to explore on firm ground the interplay between electronic correlations and non-trivial topology.
Since the local moments can randomly occupy the sites in the lattice, we also consider  the effect of disorder on the interacting topological system.

We organize the paper in the following way: In Sec. II, we
introduce the Hamiltonian, which is a generalization of the $p$-wave superconducting Kitaev chain~\cite{kitaev}, to include an interaction of the conduction electrons with a background of localized electrons  through  a Falikov-Kimball Hamiltonian~\cite{brydon}. In Sec. III, we present the formalism and the exact solution of the model. In Sec. IV, we briefly discuss the properties of the pure  one-dimensional spinless FK model, since this will be useful to characterize the superconducting phases.    Sec. V  treats the general case of a conduction band with a $p$-wave  pairing that interacts with localized $f$-electrons. We obtain the dispersion relations and  density of states of infinite chains. For a full characterization of the topological properties we investigate numerically the appearance of Majorana zero modes (MZMs) in finite chains with open boundary conditions, including the case the local moments are randomly distributed. In Sec. VI, we study in detail random chains, with different probabilities of occupations of the sites by local moments. This allow to examine the effects of disorder in the topological properties. Finally, in Sec. VII  we conclude with a summary of our results and a discussion on new perspectives and extensions of the model.

\section{The Hamiltonian}

The Hamiltonian that describes a chain of spinless fermions, with $p$-wave pairing,  interacting with a background of localized electrons through a Falicov-Kimball interaction~\cite{brydon}, can be written as a sum of three contributions. The first  is the Kitaev one~\cite{alicea,kitaev},
\begin{equation}
\mathcal{H_K}= - \sum_{i,j} t_{ij} c^{\dagger}_i c_j - \sum_{i,j}\left(\Delta_{ij} c^{\dagger}_i c^{\dagger}_j +\Delta^*_{ij} c_ic_j \right)-\mu \sum_i c^{\dagger}_ic_i.
\end{equation}

The  localized electrons~\cite{brydon} are described by,
\begin{equation}
\mathcal{H_L}= \sum_i (E_f - \mu)  f^{\dagger}_if_i,
\end{equation}
where in these equations, $t_{ij}$ is the hopping term of the conduction $c$-electrons, $\Delta_{ij}$  the $p$-wave pairing of electrons in neighboring sites, $E_f$ the energy of the localized $f$-electrons and $\mu$ the chemical potential.  Finally, the coupling between the itinerant $c$-electrons and the localized $f$-electrons is of the FK type~\cite{brydon} and given by,
\begin{equation}
\mathcal{H_{FK}}= J\sum_i  \left(f^{\dagger}_if_i - \frac{1}{2}\right) \left(c^{\dagger}_ic_i -\frac{1}{2}\right)=J \sum_i \left(n^f_i-\frac{1}{2}\right)\left(n^c_i -\frac{1}{2}\right).
\end{equation}

In order to solve the full  Hamiltonian $\mathcal{H}=\mathcal{H_K}+ \mathcal{H_L} +\mathcal{H_{FK}}$, we introduce sets of Majorana operators~\cite{shen}, $\alpha_A$, $\alpha_B$, $\beta_A$ and $\beta_B$, such that,
\begin{eqnarray} 
c_i=(\alpha_{Bi}+i \alpha_{Ai})/2, \\ \nonumber
c^{\dagger}_i=(\alpha_{Bi}- i \alpha_{Ai})/2,
\end{eqnarray}
where the Majoranas are their own anti-particles, i. e., $\alpha_{A,B}^{\dagger}=\alpha_{A,B}$.
Similarly, for the $f$-electrons we have,
\begin{eqnarray}
f_i=(\beta_{Bi}+i \beta_{Ai})/2, \\ \nonumber
f^{\dagger}_i=(\beta_{Bi}- i \beta_{Ai})/2,
\end{eqnarray}
where $\beta_{A,B}^{\dagger}=\beta_{A,B}$. In terms of these new operators we can write the interaction term as,
\begin{equation}
\mathcal{H_{FK}}= \frac{J}{4} \sum_i  (i \beta_{Bi}\beta_{Ai})(i\alpha_{Ai}\alpha_{Bi}).
\end{equation}

For the other terms in the Hamiltonian, we get in the Majorana basis,
\begin{equation}
\mathcal{H_K}+\mathcal{H_L}= -\frac{\mu}{2}\sum_i (1+\alpha_{Bi}\alpha_{Ai})-\frac{i}{2}\sum_i \left[ (\Delta+t)\alpha_{Bi}\alpha_{A i+1}+(\Delta-t)\alpha_{Ai}\alpha_{B i+1} \right ]-\frac{\mu-E_f}{2}\sum_i (1+\beta_{Bi}\beta_{Ai}).
\label{eq2.1}
\end{equation}

The many-body problem given by $\mathcal{H}=\mathcal{H_K}+ \mathcal{H_L} +\mathcal{H_{FK}}$
is complex to solve, however  a solution can be obtained in some cases. 

\section{The solution}

For further progress in understanding the properties of the model,  we consider that the chemical potential coincides with the energy of the localized quasi-particles, i.e., $\mu=E_f$. We also assume that  $\mu=E_f=0$, although the case $E_F=\mu \ne 0$  can also be solved. Notice that for $\mu=0$,  the superconducting chain, in the absence of the coupling to the localized electrons, is  in a nontrivial topological state, since the condition $|\mu/2t|<1$ is satisfied~\cite{alicea}. On the other hand, for $\Delta=0$ the model corresponds to the one-dimensional spinless FK model~\cite{brydon}.

 When $E_f=\mu=0$,  the Hamiltonian $\mathcal{H_{KL}}=\mathcal{H_{K}}+\mathcal{H_{L}}$ does not depend on the Majorana operators $\beta_B$ and $\beta_A$, associated with the localized electrons. We can then define an operator $K_i= i \beta_{Ai}\beta_{Bi}$ that  commutes with the Hamiltonian of the superconductor, i.e., $[K_i,\mathcal{H_{KL}}]=0$ for all sites. This implies that $K_i$ is a $c$-number~\cite{he}. On the other hand, the anti-commutation rules of the Majorana operators imply that $K_i^2=1$ or $K_i=\pm 1$. The Hamiltonian of the system can now be written as,
\begin{equation}
\label{HT}
\mathcal{H}= -\frac{i}{2}\sum_i \left[ (\Delta+t)\alpha_{Bi}\alpha_{A i+1}+(\Delta-t)\alpha_{Ai}\alpha_{B i+1} \right ]+ i \sum_i   J_i \alpha_{Ai}\alpha_{Bi},
\end{equation}
where $J_i=J K_i/16$ ($K_i=\pm 1$). From a formal point of view $\mathcal{H}$ is now a quadratic form that can be exactly solved. The interacting problem has been mapped in a non-interacting one, formally in a system with a site dependent chemical potential~\cite{gotta1,gotta2,gotta3}.

Let us consider the case of an infinite chain and introduce the quantities,
\begin{equation}
\alpha_{Ai} =\sum_k e^{ikr_i} \alpha_A(k),
\end{equation}
and an equivalent equation for $\alpha_{Bi}$. Since $\alpha_A^{\dagger}(k)=\alpha_A(-k)$, only $\alpha_A(k=0)$ is a Majorana operator~\cite{shen}.
In terms of these operators the Hamiltonian can be written as,
\begin{equation}
\label{eq}
\mathcal{H}= -\frac{i}{2}\sum_k \left[ (\Delta+t)e^{ik} \alpha_B(k)\alpha_A(-k) - (\Delta-t)e^{-ik}\alpha_B(k) \alpha_A(-k)  \right]+ i J \sum_k\alpha_A(k) \alpha_B(-(k+Q)),
\end{equation}
where we wrote $J_i=J e^{i Q r_i}$.
The homogeneous configuration, with a localized electron on each site of the lattice, $n_{fi}=1$,  corresponds to $Q=0$. The staggered configuration,  where the values of $K_i$ alternate between $\pm1$, such that $n_{fi}$ takes values $0$ and $1$ in consecutive sites, corresponds to $Q= \pi$. Notice that $n_{fi}=(1/2)(1+i \beta_{Ai}\beta_{Bi})$.   Among all possible configurations, these two play an important role,  as we  discuss below. 

\begin{figure}[tbh]
\centering
\includegraphics[width=0.95\columnwidth]{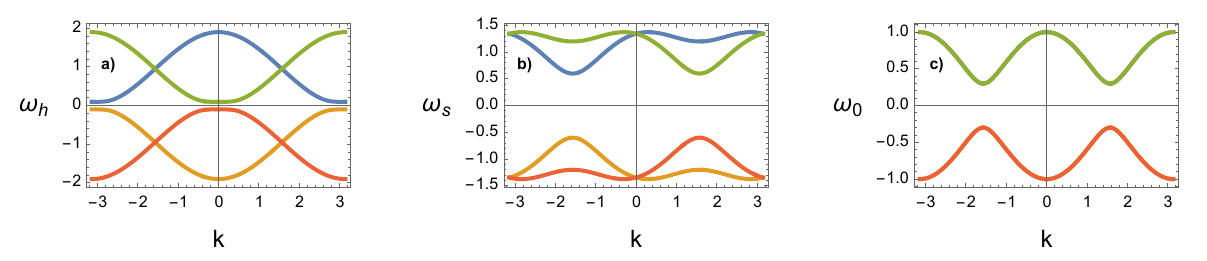}
\caption{(Color online) Dispersion relations of the excitations in the superconducting state with $t=1$, $\Delta/t=0.2$ and $J/t=0.9$. a) homogeneous case ($Q=0$, $\langle n_f  \rangle=1$), $\omega_h(k)$, and b) staggered case ($Q=\pi$, $\langle n_f \rangle=1/2$), $\omega_s(k)$. c)  energy of the bogoliubons $\omega_0(k)$, of the non-interacting case with the same parameters, but  $J=0$ (Kitaev chain). The conduction band is always half-filled, and $\mu=E_f=0$. }
\label{fig1}
\end{figure}

The excitations of the system are given by the k-dependent eigenvalues of the Hamiltonian Eq.~\ref{eq}~\cite{alicea,altland}.  For the homogeneous configuration ($Q=0$), we get,
\begin{equation}
\label{drw}
\omega_{h(1,2)}(k)=\frac{\sqrt{\Delta ^2+2 J^2 \pm 4 J t \cos (k)+\cos (2 k) \left(t^2-\Delta ^2\right)+t^2}}{\sqrt{2}},
\end{equation}
and $\omega_{h(3,4)}(k)=\pm \omega_{h(1,2)}(k)$.
The excitations in the staggered configuration, with $Q=\pi$, are given by
\begin{equation}
\label{drs}
\omega_{s(1,2)}(k)=\frac{\sqrt{\Delta ^2+2 J^2 \pm 4 \Delta  J \sin (k)-\cos (2 k) \left(t^2-\Delta ^2\right)+t^2}}{\sqrt{2}},
\end{equation}
and $\omega_{s(3,4)} (k)=\pm \omega_{s(2,1)}(k)$.

Notice that the condition $\mu=E_F=0$ implies that for the non-interacting system, the band of conduction electrons is  half-filled. The number of localized $f$-electrons is arbitrary,  $n_f \le 1$, and does not affect the total energy of the system, when  $J=0$.

In Fig.~\ref{fig1} we plot the dispersion relations of the full problem, Eqs.~\ref{drw} and~\ref{drs}, for specific values of parameters $\Delta$ and $J$, of the homogeneous and staggered configurations. For completeness, we also show the dispersions of the pure Kitaev chain ($J=0$). In the homogeneous case of Fig.~\ref{fig1}a, the gap closes for $J/t=1$, independent of $\Delta$. In the staggered case, Fig.~\ref{fig1}b, the gap in the dispersions closes for $J=\Delta$.

In Section IV, to fully describe the topological properties of the interacting system, we  investigate  numerically the case of finite chains with open boundary conditions.   We obtain the spectrum of eigenvalues and the wave functions of the MZMs. However, before that  we discuss briefly the pure one-dimensional spinless FK model, since its properties will be relevant for understanding the superconducting case. Notice that this model corresponds to taking the superconducting gap $\Delta=0$ in Eq.~\ref{HT}.

\section{The one-dimensional spinless Falicov-Kimball model ($\Delta=0$)}

The FK model is a well studied strongly correlated electronic problem, with many applications ranging from heavy fermions to metal-insulator and valence transitions~\cite{brydon}. In particular the one-dimensional spinless FK model has many well-known properties. It can be approached using the technique of bosonisation that yields exact results.  From the point of view of topological properties this model is completely trivial.

In this section we briefly  discuss the spinless one-dimensional FK model and emphasize some aspects that are relevant when considering the full superconducting problem. Formally, we just take the amplitude of the superconducting pairing $\Delta=0$, to obtain the dispersion relations for the homogeneous and staggered phases of the model. It is possible to  distinguish two situations depending on the value of the ratio $(J/t)$.
\begin{figure}[tbh]
\begin{subfigure}[h]{0.33\linewidth}
\includegraphics[width=\linewidth]{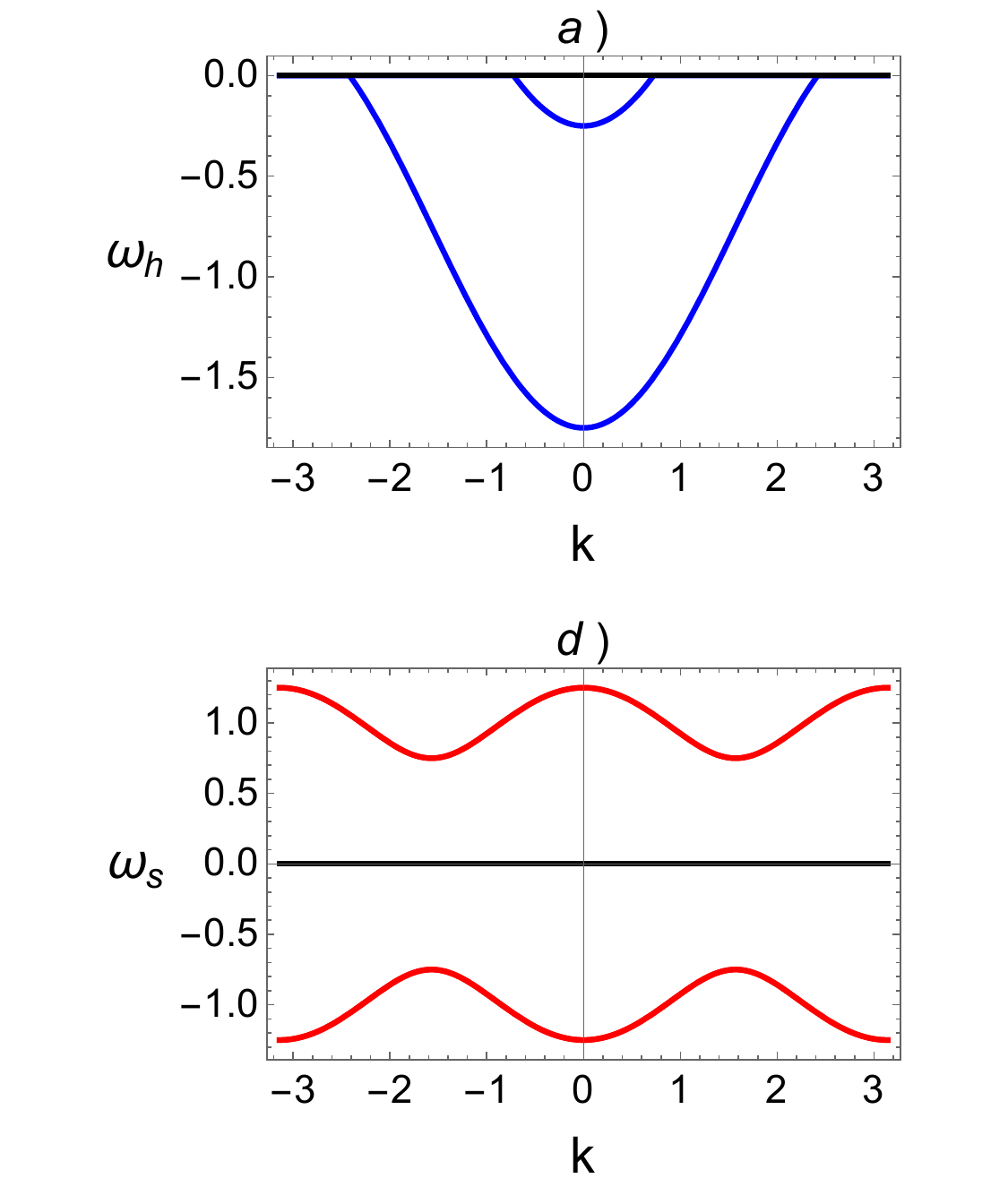}
\end{subfigure}%
\begin{subfigure}[h]{0.33\linewidth}
\includegraphics[width=\linewidth]{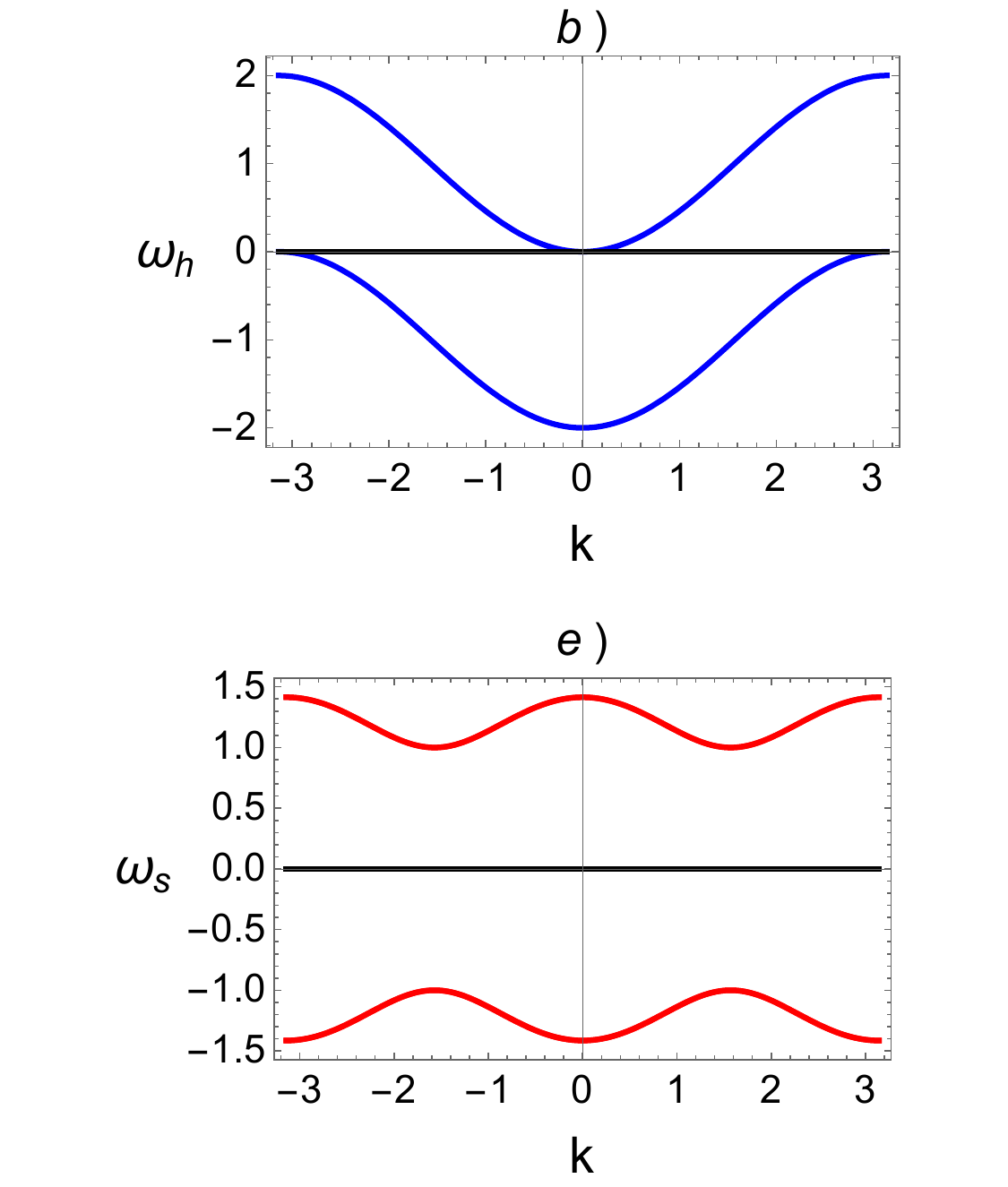}
\end{subfigure}
\begin{subfigure}[h]{0.33\linewidth}
\includegraphics[width=\linewidth]{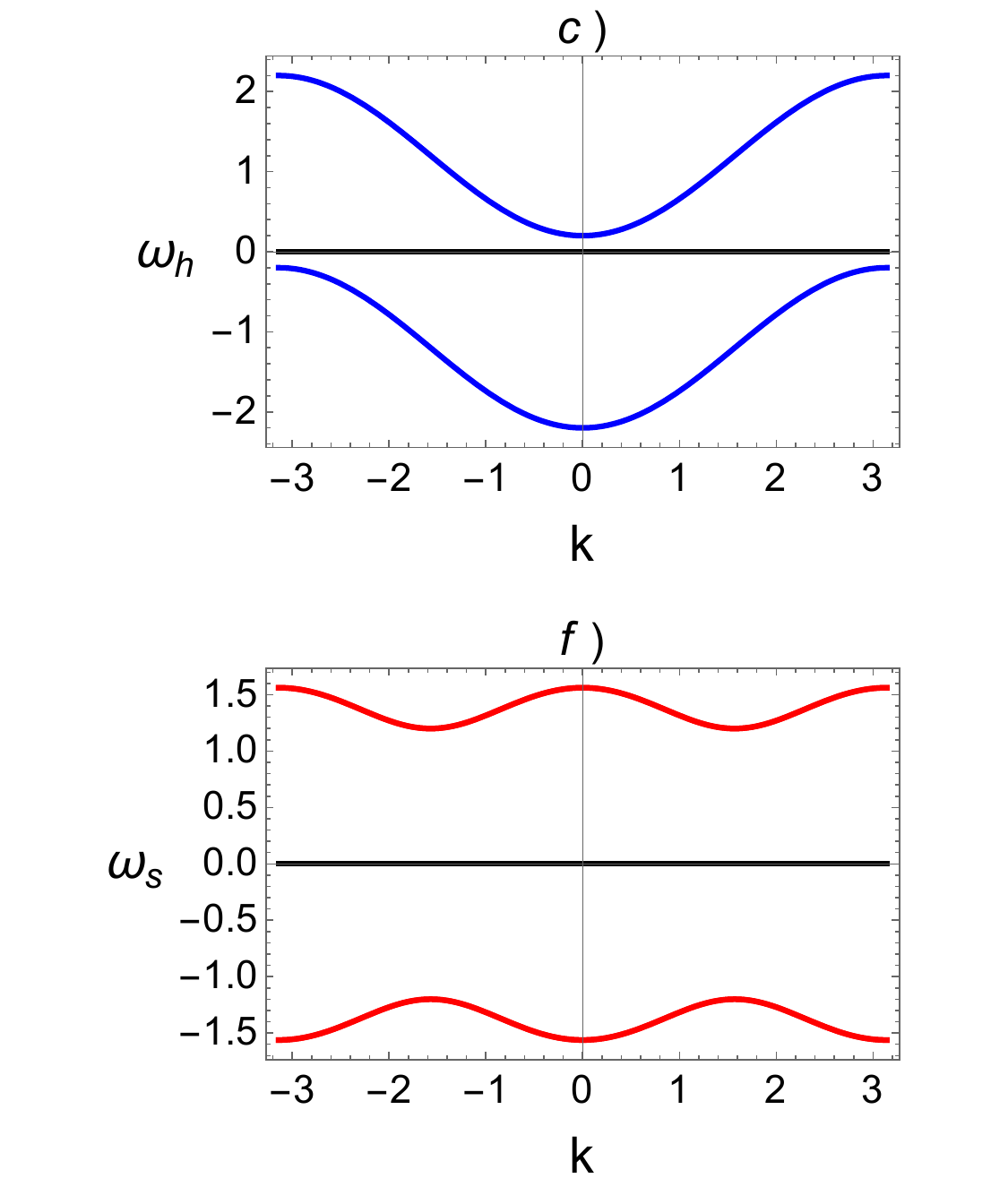}
\end{subfigure}%
\caption{Dispersion relations of the  FK model ($\Delta=0$). Homogeneous case, $\omega_h(k)$   for a) $J/t=0.75$, b) $J/t=1$ and c) $J/t=1.2$. Staggered configuration, $\omega_s(k)$ for d) $J/t=0.75$, e) $J/t=1$ and f) $J/t=1.2$.  In the homogeneous  case,  a) corresponds to a metallic state and at b) $J/t=1$ there is a metal insulator transition. In the staggered configuration the system is always insulating. The black line is the energy of the localized quasiparticles that coincides with the chemical potential. }
\label{fig2}
\end{figure}
For $J/t>1$, both homogeneous and staggered configurations correspond to insulating phases, as shown in Figs.~\ref{fig2}c and~\ref{fig2}f. Conversely,  for $J/t <1$,  the homogeneous configuration corresponds to a metallic phase, Fig.~\ref{fig2}a, but the staggered one remains insulating, as can be seen in Fig.~\ref{fig2}d.   
Then for the homogeneous case, with one localized electron per site ($n_f=1$), there is a metal-insulator transition at $|J/t|=1$, as shown in Fig.~\ref{fig2}b for the dispersion relations. The staggered phase is always insulating. Notice that the presence of two dispersive bands is due to an effective hybridization between the localized level and the conduction band due to the FK interaction. If this is too large it can open a gap and give rise to a metal-insulator transition

\begin{figure}[ht]
\centering
\includegraphics[width=0.35\columnwidth]{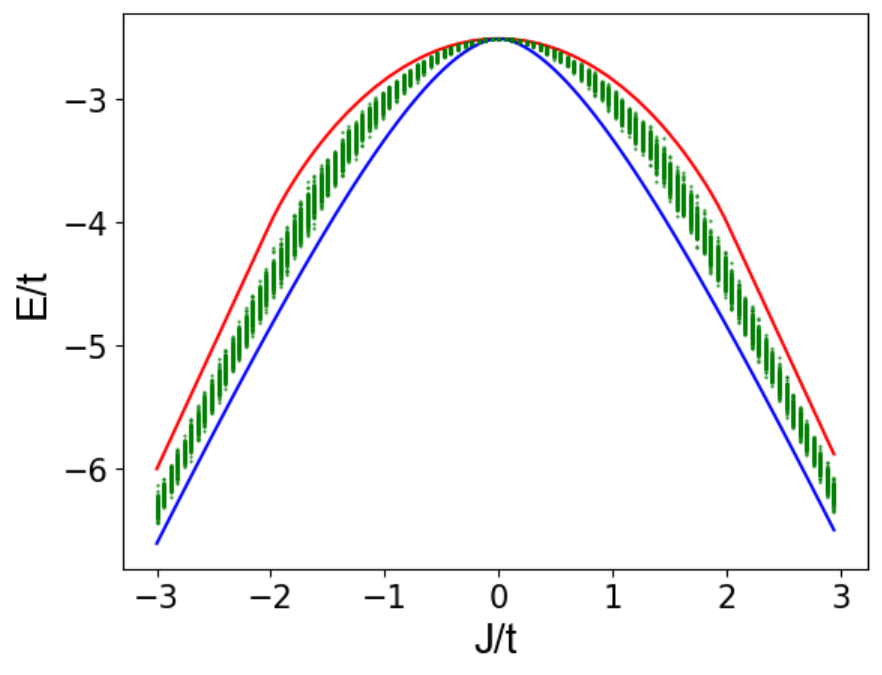}
\caption{(Color online) Ground state energy of the spinless FK model ($\Delta=0$). Homogeneous case, $n_f=1$ (red),   staggered, $n_f=1/2$  (blue) and  random configurations (green) with $\langle n_f \rangle=1/2$.  Notice that, the staggered configuration has an energy lower than all the probed random configurations with $\langle n_f \rangle=1/2$. }
\label{fig3}
\end{figure}

Due to the condition $E_f=\mu=0$, the energy of the pure $f$-electron system does not depend on the number of $f$-electrons, differently from the coupled $s-f$ system. In Fig.~\ref{fig3} we show the ground state energies of the special configurations, $Q=0$ and $Q=\pi$, as  functions of the interaction $J$. Notice that the staggered configuration has a lower energy compared to the homogeneous one.

When $\langle n_f \rangle=1/2$, besides ordered configurations, as the staggered and other types of superlattices~\cite{gotta3}, there are many disordered configurations~\cite{brydon, molecule}, corresponding to a random arrangement of the local moments in the chain.  
In Fig.~\ref{fig3} we also show the ground state energies,  as functions of the interaction $J$, for several realizations of random configurations with $\langle n_f \rangle=1/2$.  The disordered configurations have a probability $p=1/2$ for a site being occupied,  by a localized $f$-electron. The ordered,  staggered configuration is always the one with lower energy. 

\section{Topological superconductivity in the presence of Interaction between itinerant and localized electrons}

\subsection{Analysis of the dispersion relations and density of states}

We now return to the full problem consisting of a band of spinless fermions, with  $p$-wave pairing,  that interacts with a background of localized $f$-electrons, Eq.~\ref{HT}.  The Hamiltonian with $J=0$, corresponds to a Kitaev chain and an independent collection of localized electrons. On the other hand, for $\Delta=0$, we obtain the FK model.

The topological properties of the Kitaev chain are well known, it presents a non-trivial topological phase, for $|\mu/2t|<1$, with Majorana modes at its ends~\cite{alicea}. In the present case the chemical potential $\mu=0$, such that the Kitaev chain is in a topological superconducting phase~\cite{alicea}.  We now turn on the interaction with the localized $f$-electrons and study its effects.

\begin{figure}[H]
\centering
\begin{subfigure}[h]{0.36\linewidth}
\includegraphics[width=\linewidth]{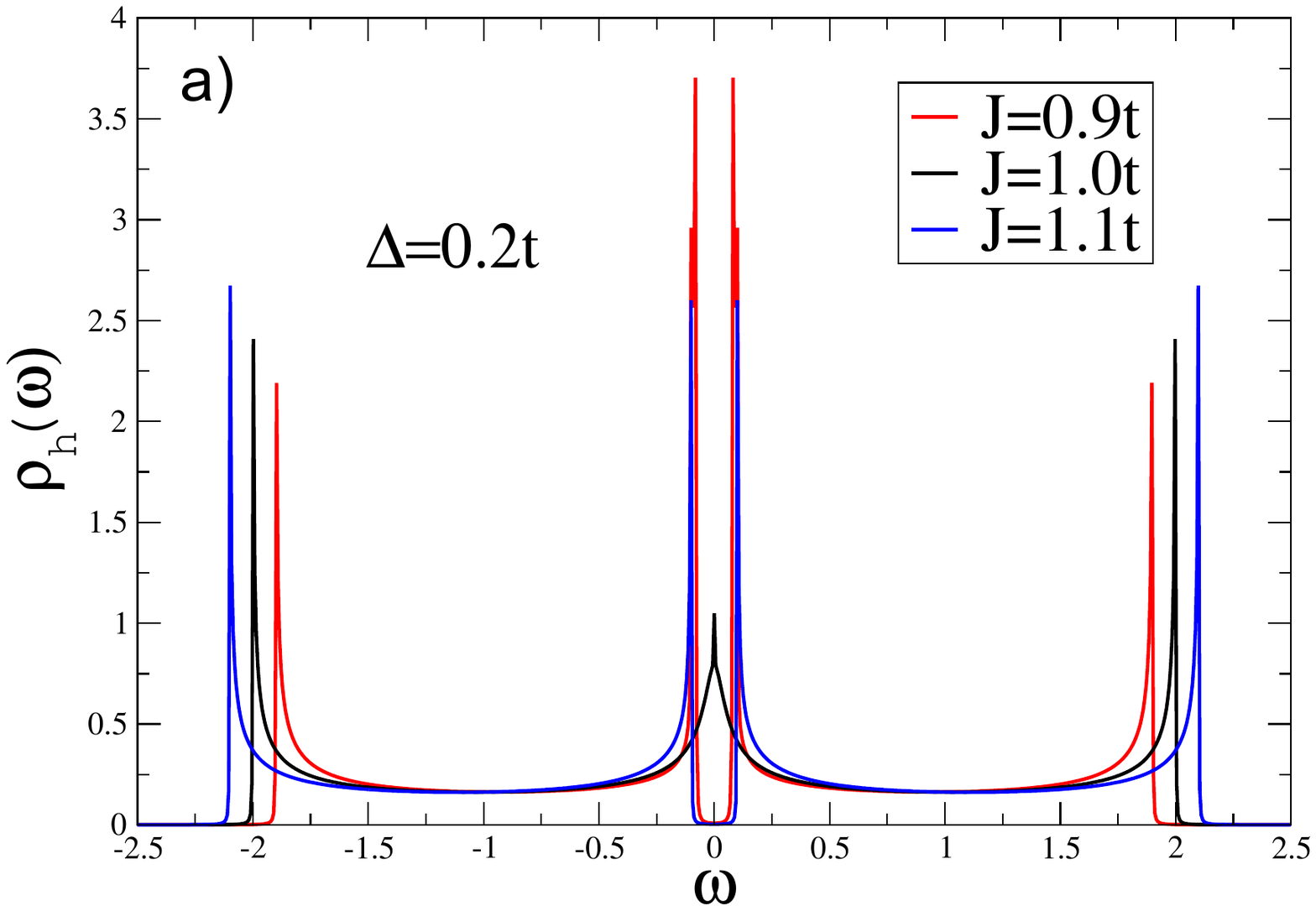}
\end{subfigure}

\begin{subfigure}[h]{0.36\linewidth}
\includegraphics[width=\linewidth]{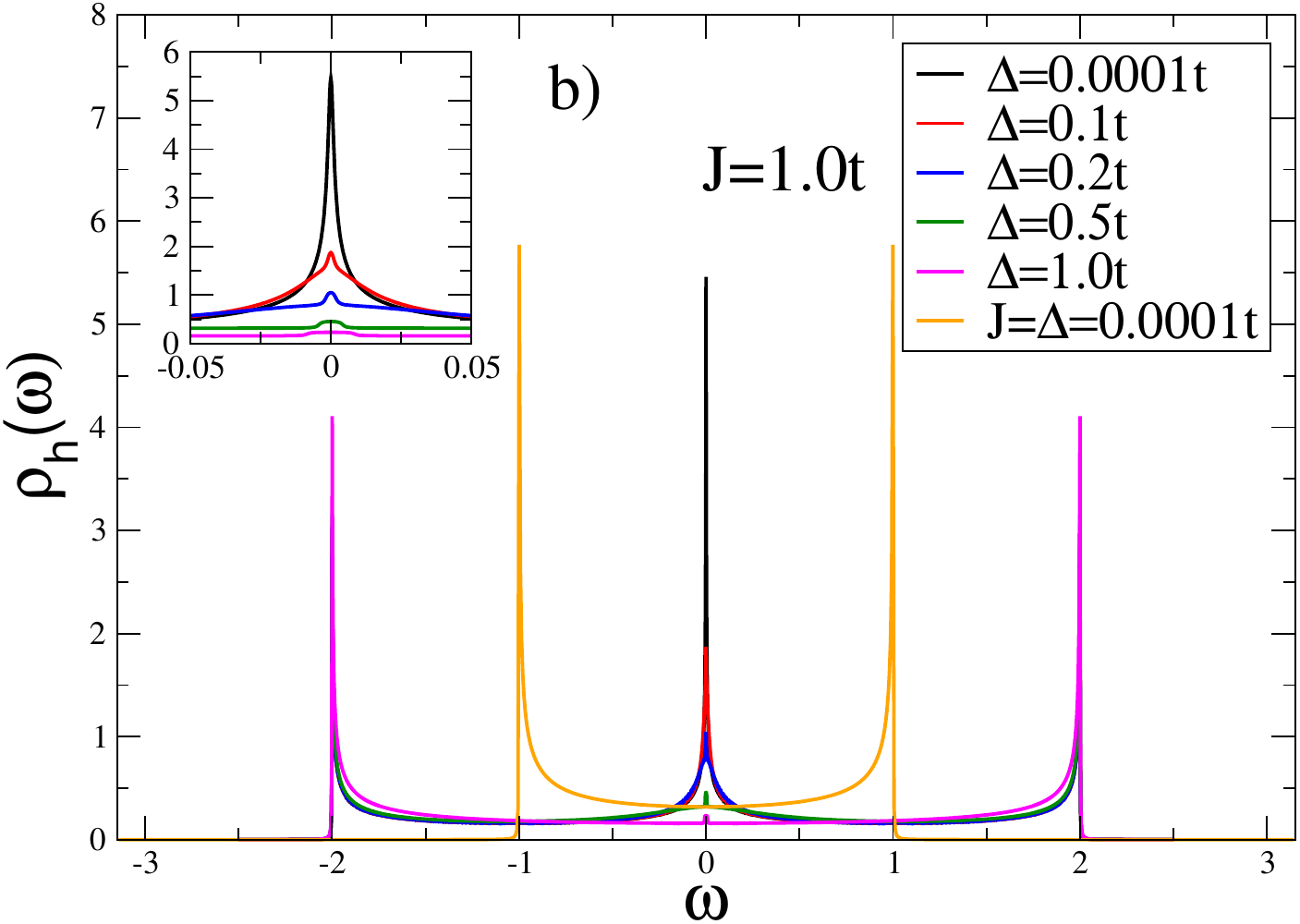}
\end{subfigure}

\begin{subfigure}[h]{0.36\linewidth}
\includegraphics[width=\linewidth]{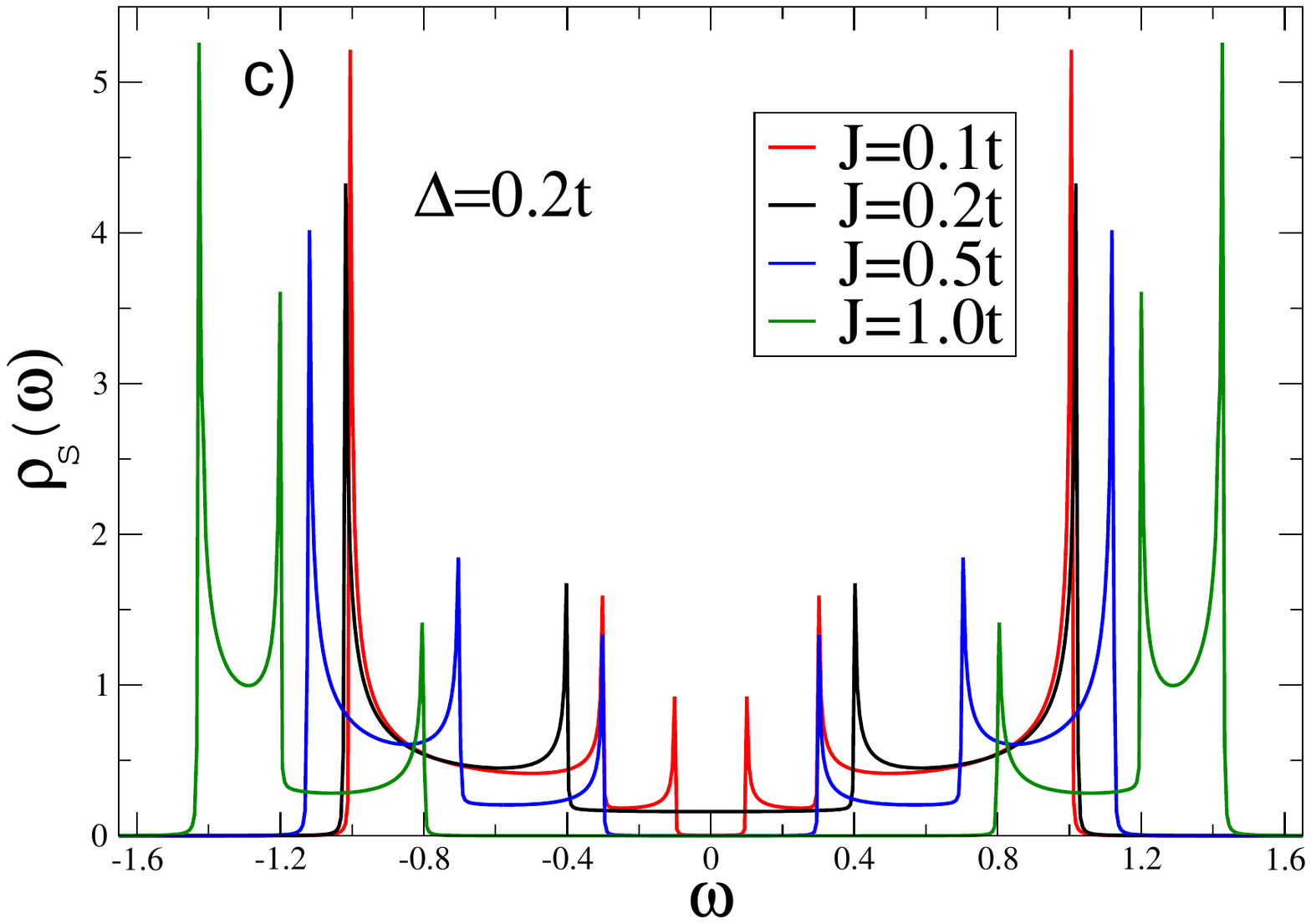}
\end{subfigure}
\caption{Density of states in the superconducting state. a) for the homogeneous case with $\Delta/t=0.2$ and different values of the coupling parameter $J/t$, below, at, and above the gap closing transition.     b)  for the homogeneous  case at the gap closing transition and different pairing amplitudes $\Delta/t$. The inset shows the density of states close to $\omega=0$. c) for the staggered configuration as a function of $J/t$ and  fixed $\Delta/t=0.2$.  }
\label{fig4}
\end{figure}

We start discussing the results of Fig.~\ref{fig1} for the dispersion relations, Eqs.~\ref{drw} and~\ref{drs}, and the density of states obtained from these dispersions shown in Fig.~\ref{fig4}. We consider both, the homogeneous and staggered configurations and take $\Delta/t=0.2$ in the figures. In the homogeneous case,  the gap closes at $J/t=1$, as can be seen in the density of states shown in Fig.~\ref{fig4}a.   For $J\ne 1$ the superconductor is gapped, as can be obtained directly from the dispersion relations and as shown in the density of states in Fig.~\ref{fig4}a. However, there is a fundamental difference between the cases $J/t<1$ and $J/t>1$, as we discuss below using numerical results in finite chains. 
Fig.~\ref{fig4}b  shows the density of states at the gap closing point of the homogeneous case for different values of the coupling $\Delta$.  As the  pairing interaction  is turned on, a competition between the localization due to FK interactions and the formation of $p$-wave  pairs occurs; the peak decreases as the superconducting pairs are formed, as indicated in the inset of the figure, and the density of states approaches that of a renormalized uncorrelated tight-binding chain. For comparison, we also show in Fig.~\ref{fig4}b the  tight-binding density of states for $J/t=\Delta/t=0.0001t$.   

In the staggered configuration,   the gap closing occurs at $J=\Delta$ and  the superconductor is gapped, whenever $J\ne \Delta$, as shown in Fig.~\ref{fig4}c for the density of states. Notice that there is a finite density of states at  $\mu=0$ for $J=\Delta$,  where the system is gapless. This is similar to the corresponding density of states of an armchair nano-ribbon, whenever $N=3M-1$, with $M$ being integer~\cite{alase}.
Like in the homogeneous case,  in the staggered phase the gapped phases,  for $J>\Delta$ and $J< \Delta$ are distinct with respect to their topological properties, as we discuss next. 

\subsection{Topological properties of finite chains}

In order to investigate the topological properties of the interacting system, instead of calculating a topological invariant~\cite{gotta1,gotta2,gotta3} and using the bulk-boundary correspondence principle~\cite{alase}, we opt to study numerically  the behavior of interacting finite chains with open boundary conditions. In this approach the appearance of   zero energy edge states gives direct evidence of the non-trivial topological character of the system.
\begin{figure}[H]
\centering
\includegraphics[width=0.7\columnwidth]{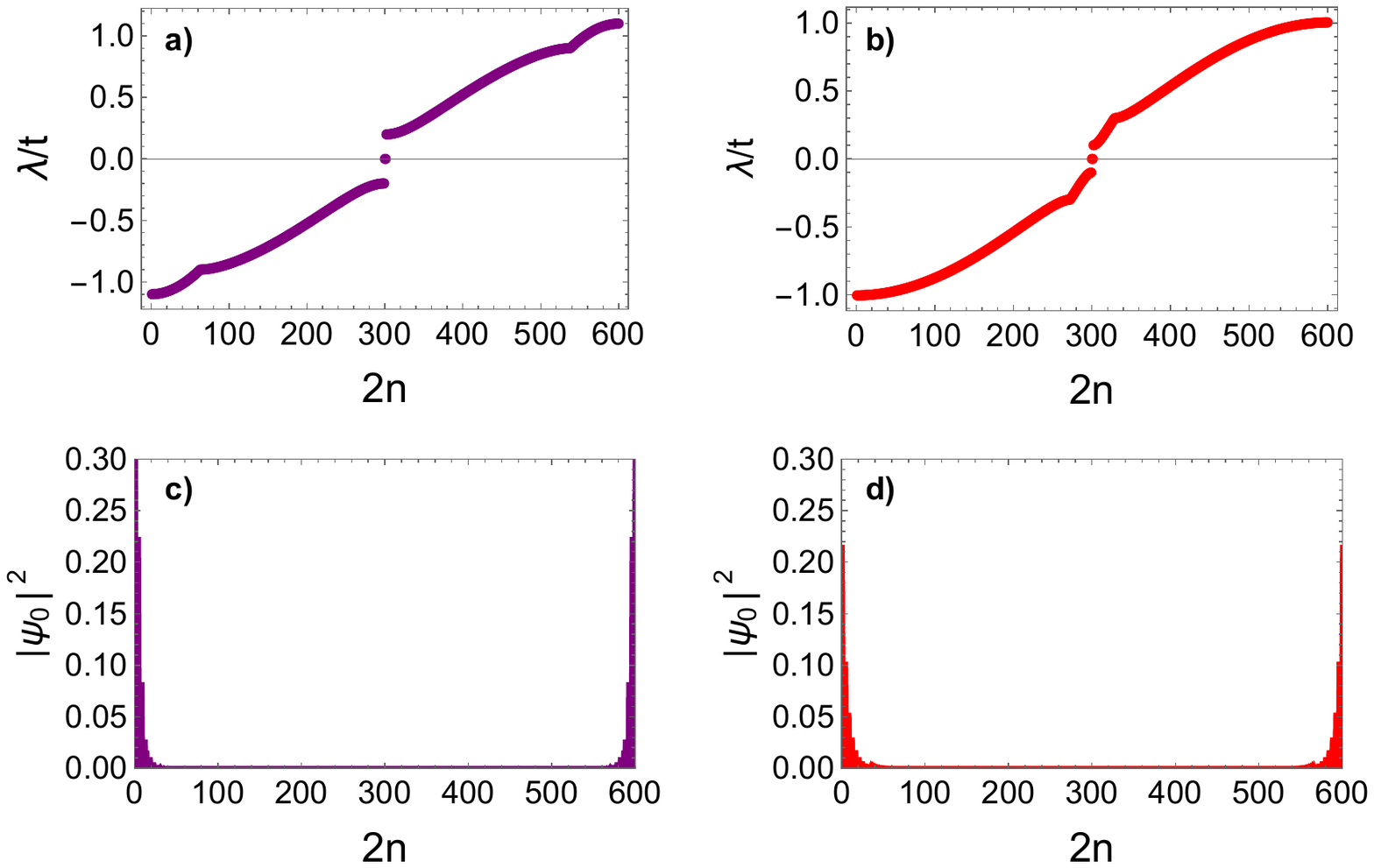}
\caption{(Color online) Eigenvalues $\lambda/t$ for a chain  of 300 sites  for $t=1$, $\Delta/t=0.2$, a) in the homogeneous phase, for $J/t=0.1<1$, b) in the staggered phase for $J/t=0.1<\Delta/t$. In  c) and d) we show  the amplitude square of the wave functions of the corresponding zero modes for the homogeneous and staggered phases,  respectively.}
\label{fig5}
\end{figure}

\begin{figure}[H]
\centering
\includegraphics[width=0.7\columnwidth]{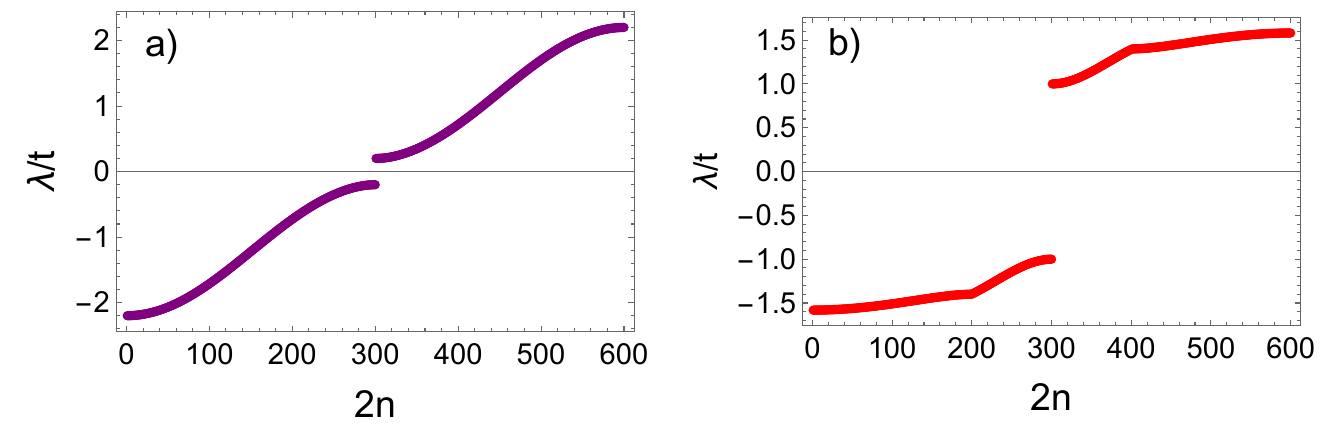}
\caption{(Color online) Eigenvalues  $\lambda/t$ for a chain  of 300 sites  for $t=1$ and fixed $\Delta/t=0.2$, a) in the homogeneous phase, for $J/t=1.3>1$, b) in the staggered phase for $J/t=1.2 >\Delta/t$. Both phases are gapped and topologically trivial superconductors as indicated by the absence of zero energy  edge  modes. }
\label{fig6}
\end{figure}

In Fig.~\ref{fig5} we show the eigenvalues for a chain of 300 sites for both homogeneous and staggered configurations. In the  homogeneous chain, Fig.~\ref{fig5}a, superconductivity  is gapped, with zero energy modes that persist, as long as $J<t$. These zero modes are edge  states as can be seen directly from their wave functions shown in Fig.~\ref{fig5}c. At $J/t=1$ the gap closes and the system is a gapless superconductor~\cite{altland}. For $J>t$, the gap reopens as shown in Fig.~\ref{fig6}a,  but the zero energy edge  states disappear,  indicating that the superconductor is topologically trivial . Recall that in the absence of pairing, the homogeneous FK chain is metallic for $J/t<1$, as shown in Fig.~\ref{fig2}a, such that topological superconductivity arises from a metallic state. Then in the homogeneous case there is a topological phase transition together with a gap closing, at $J/t=1$.

In the staggered case, the topological superconducting phase, characterized by the presence of zero energy edge  states, exists for $J<\Delta$, as shown in Fig.~\ref{fig5}b and Fig.~\ref{fig5}d. At $J=\Delta$ the superconductor is gapless and above this critical value, the superconducting chain is gapped and topologically trivial with no zero modes, as shown in Fig.~\ref{fig6}b. In the staggered case, as we turn on the superconducting pairing, non-trivial topological superconductivity   emerges from an insulating phase (see Fig.~\ref{fig2}d) of the FK model, as soon as, the pairing amplitude becomes larger than $J$. 

Then,  the topological phase transition  occurs for $J/t=1$ in the homogeneous case and for $J/\Delta=1$ in the staggered case.

\section{Random occupation of the local moments in the chain}

\begin{figure}[H]
\centering
\includegraphics[width=.35\textwidth]{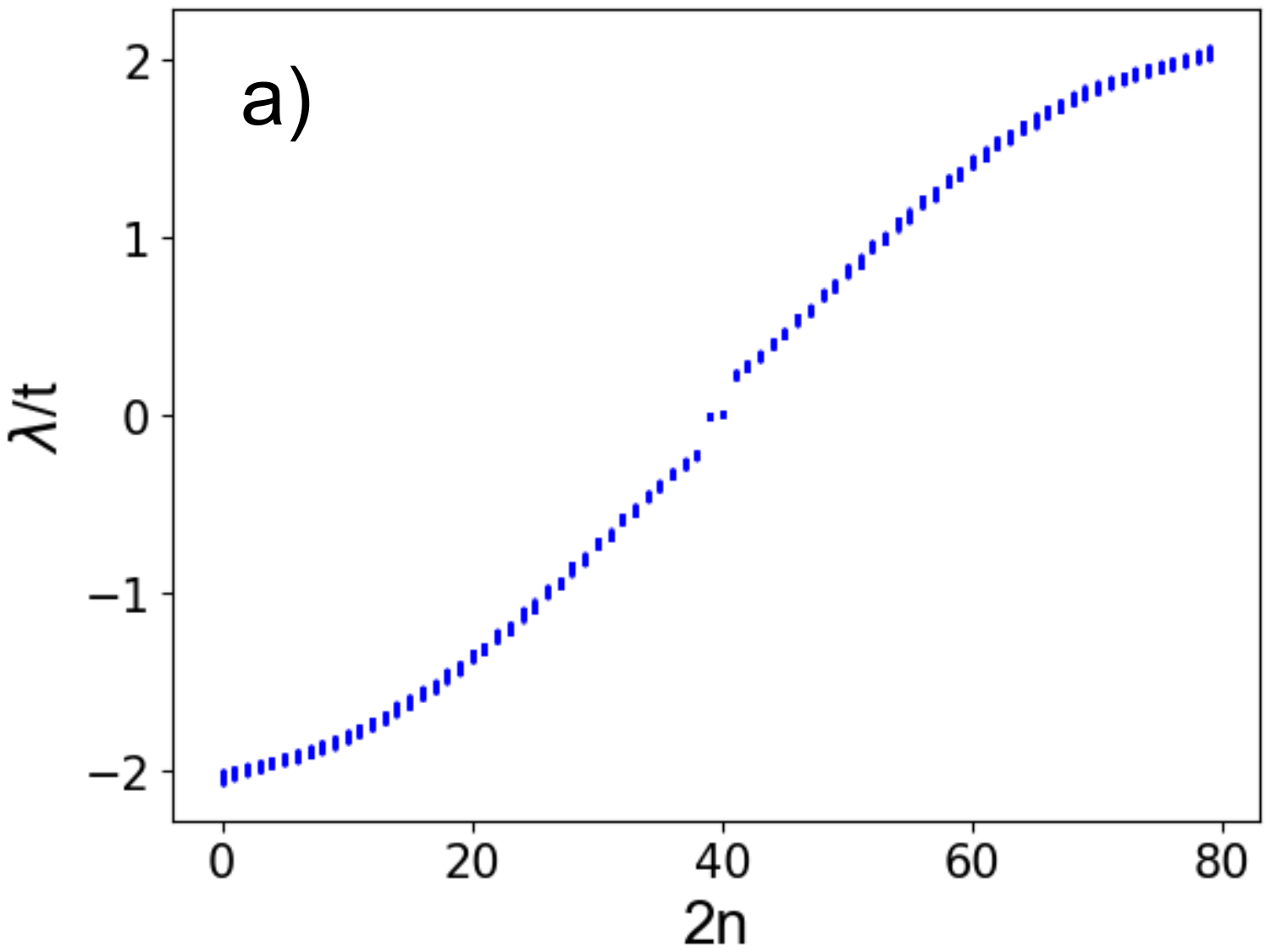}\quad
\includegraphics[width=.35\textwidth]{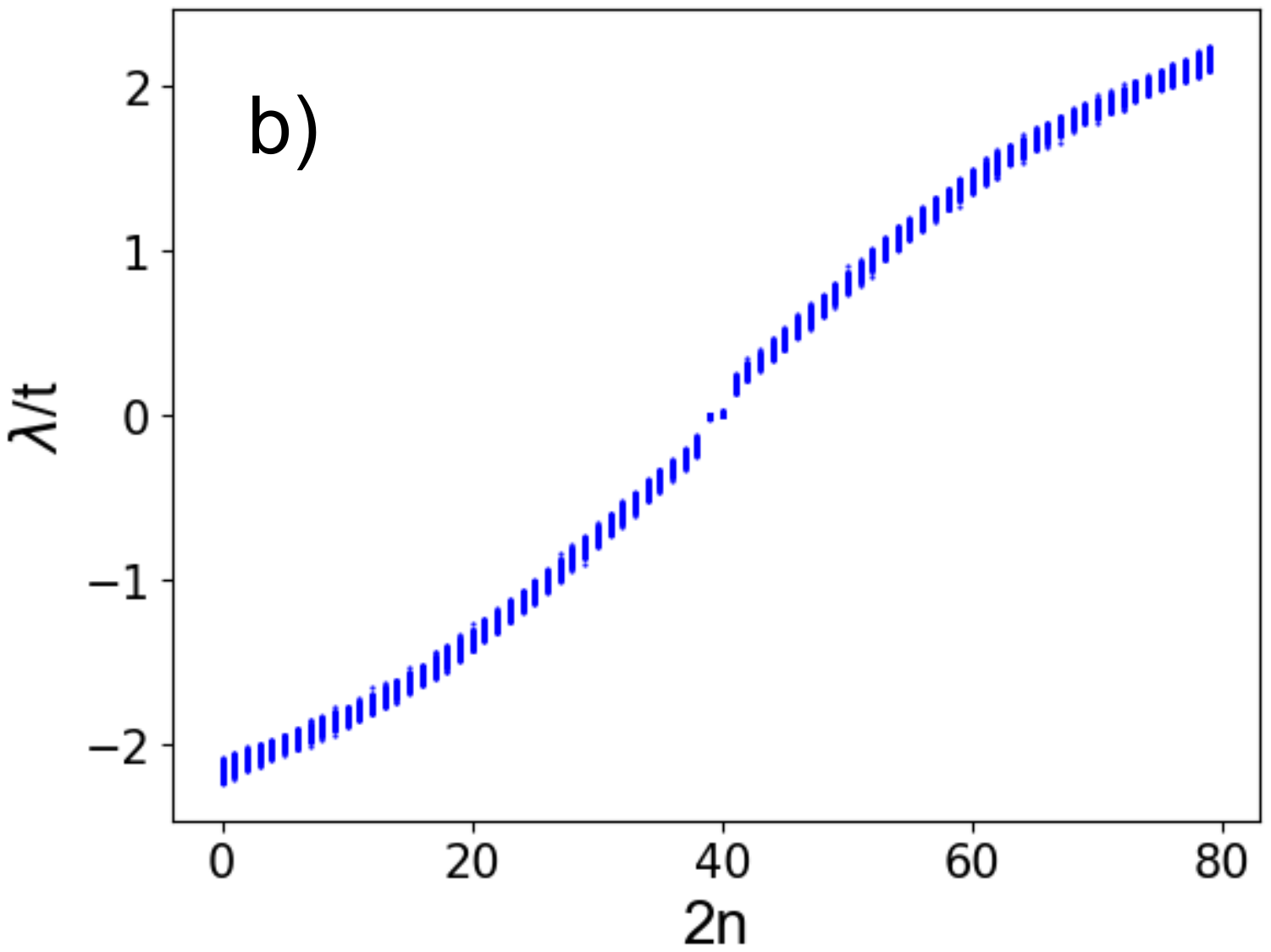}\quad
\includegraphics[width=.35\textwidth]{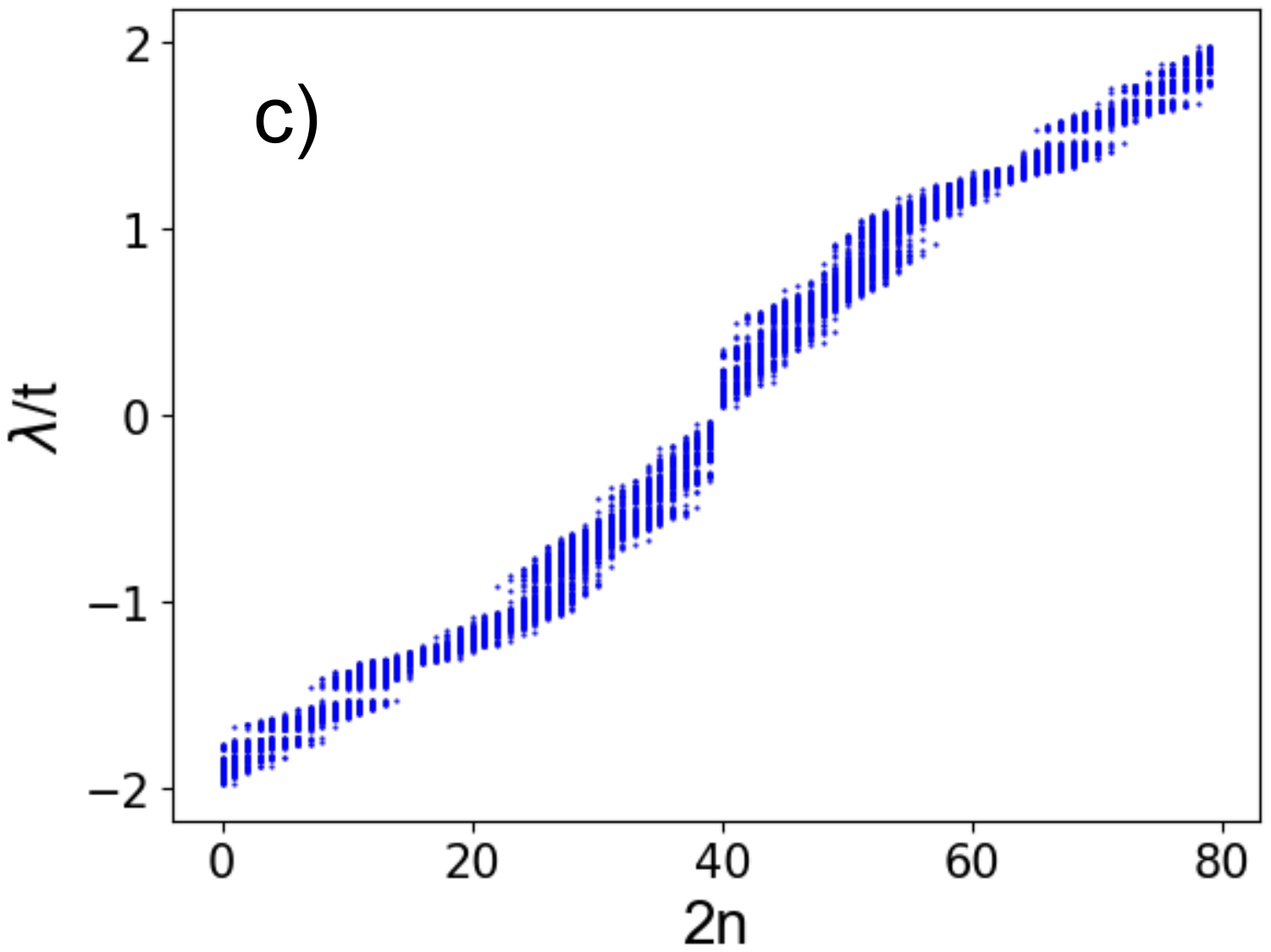}\quad
\includegraphics[width=.35\textwidth]{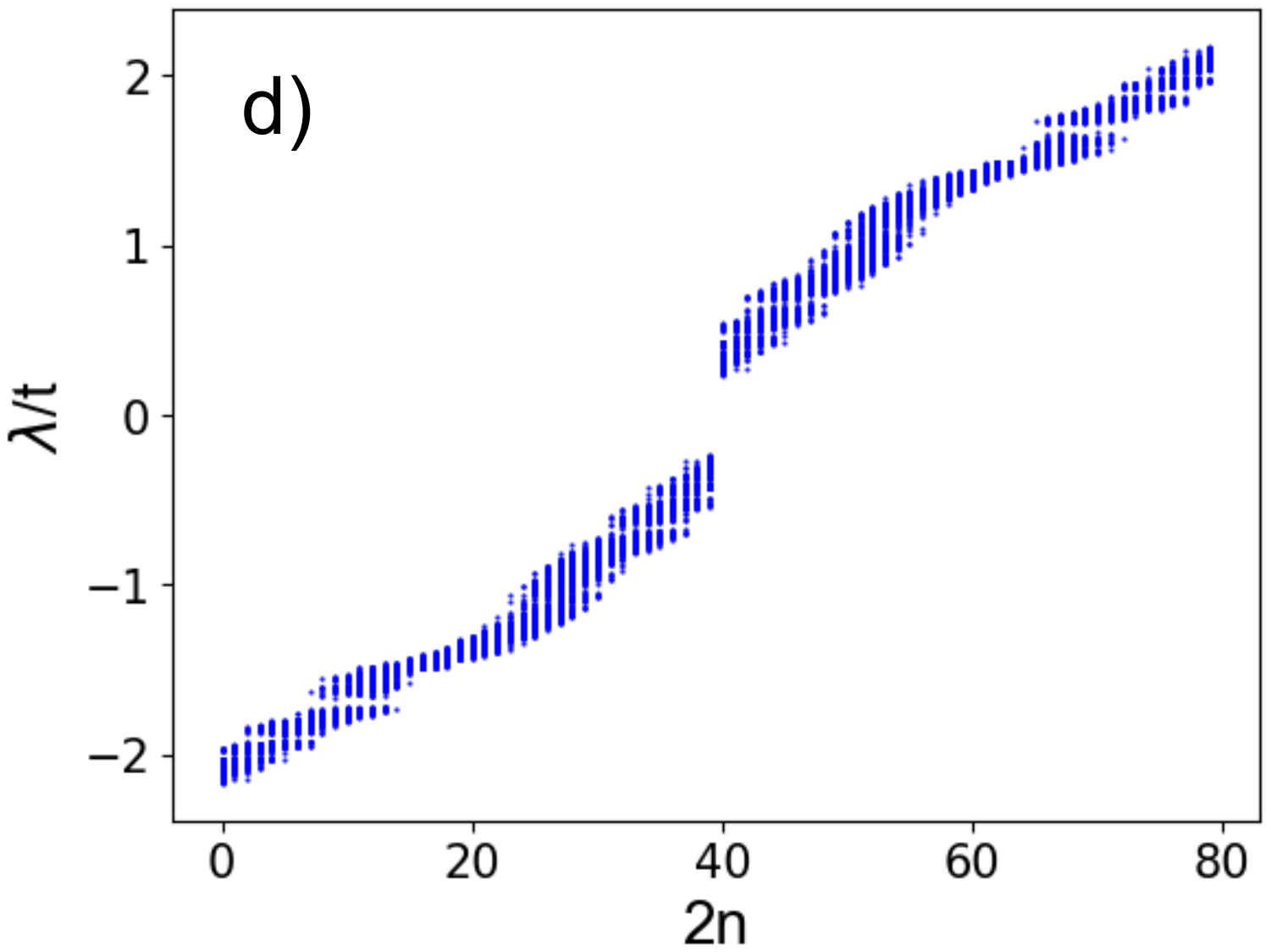}
\caption{ Eigenvalues $\lambda/t$ for $\Delta/t=0.2$, corresponding to 200 realizations of random configurations for a chain with 40 sites, with probability $p=1/2$ for a site  being occupied by a localized electron ($\langle n_f \rangle=1/2$).   a) for $J/t=0.1$, b)  for $J/t=0.3$,  c) for $J/t=1$ and d) for $J/t=1.2$  The zero modes persist all the way to the topological phase transition of the homogeneous case at $J/t=1$.}
\label{fig7}
\end{figure}

\begin{figure}[H]
\begin{subfigure}[h]{0.32\linewidth}
\includegraphics[width=\linewidth]{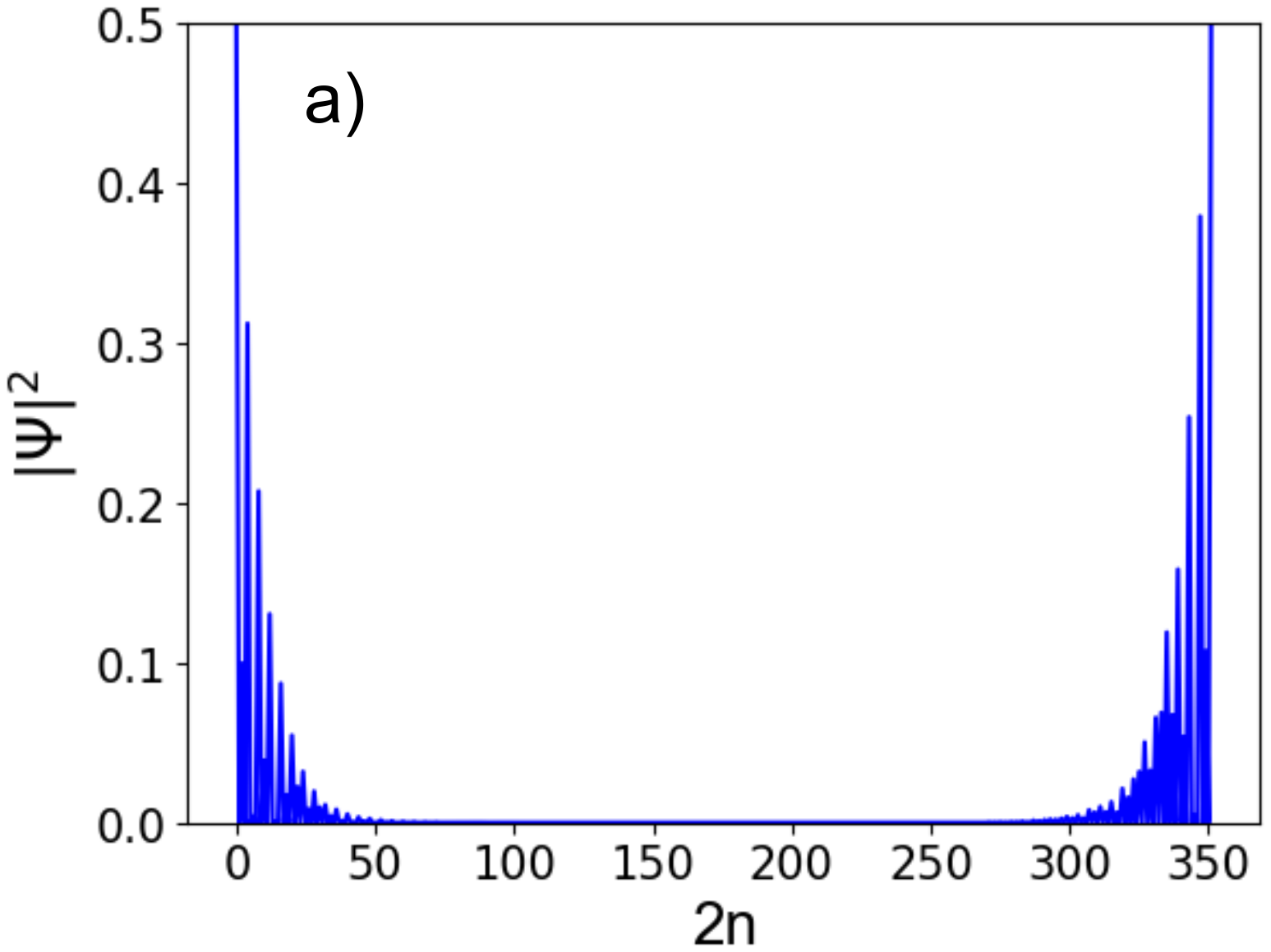}
\end{subfigure}
\begin{subfigure}[h]{0.32\linewidth}
\includegraphics[width=\linewidth]{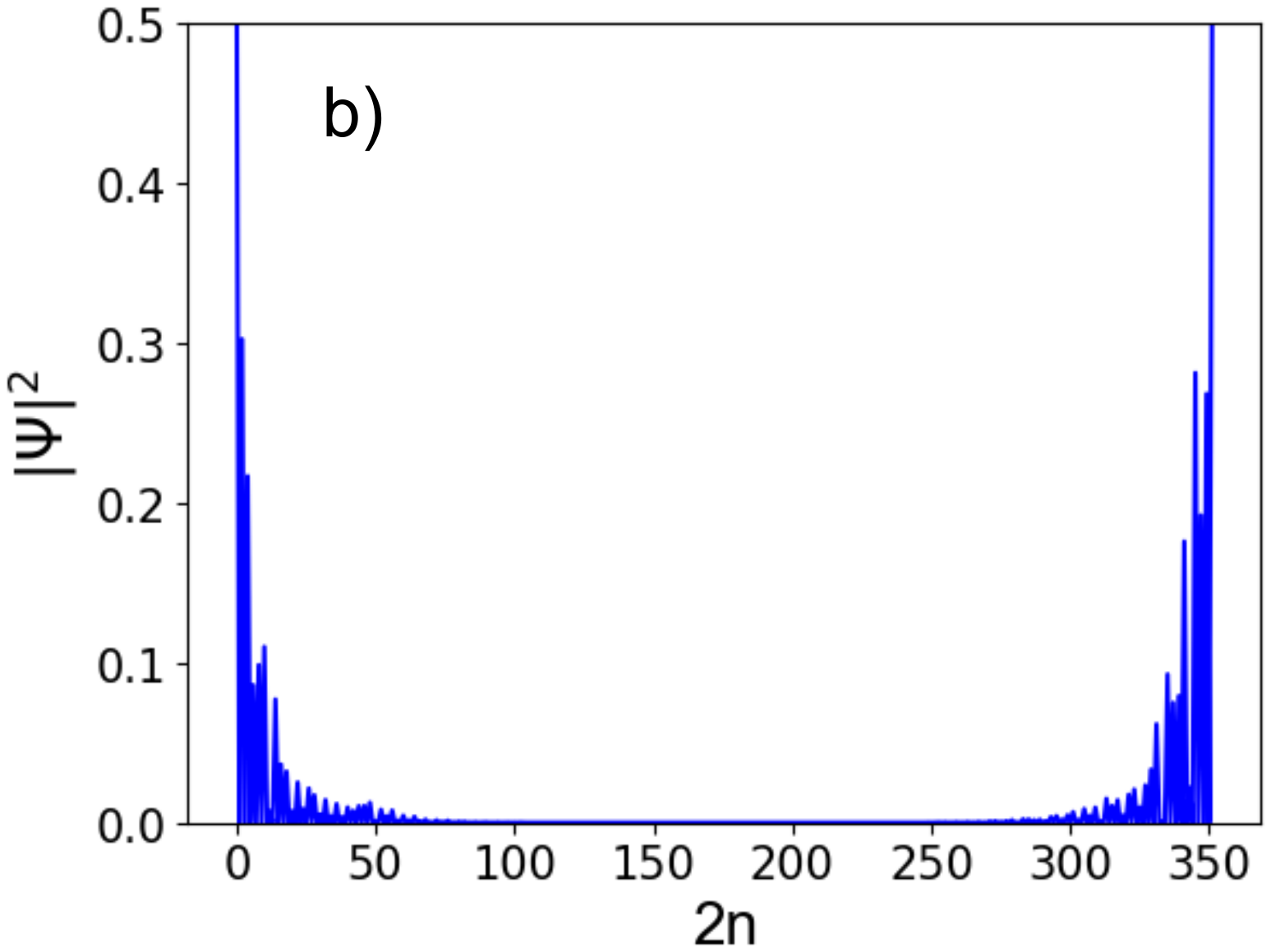}
\end{subfigure}
\begin{subfigure}[h]{0.32\linewidth}
\includegraphics[width=\linewidth]{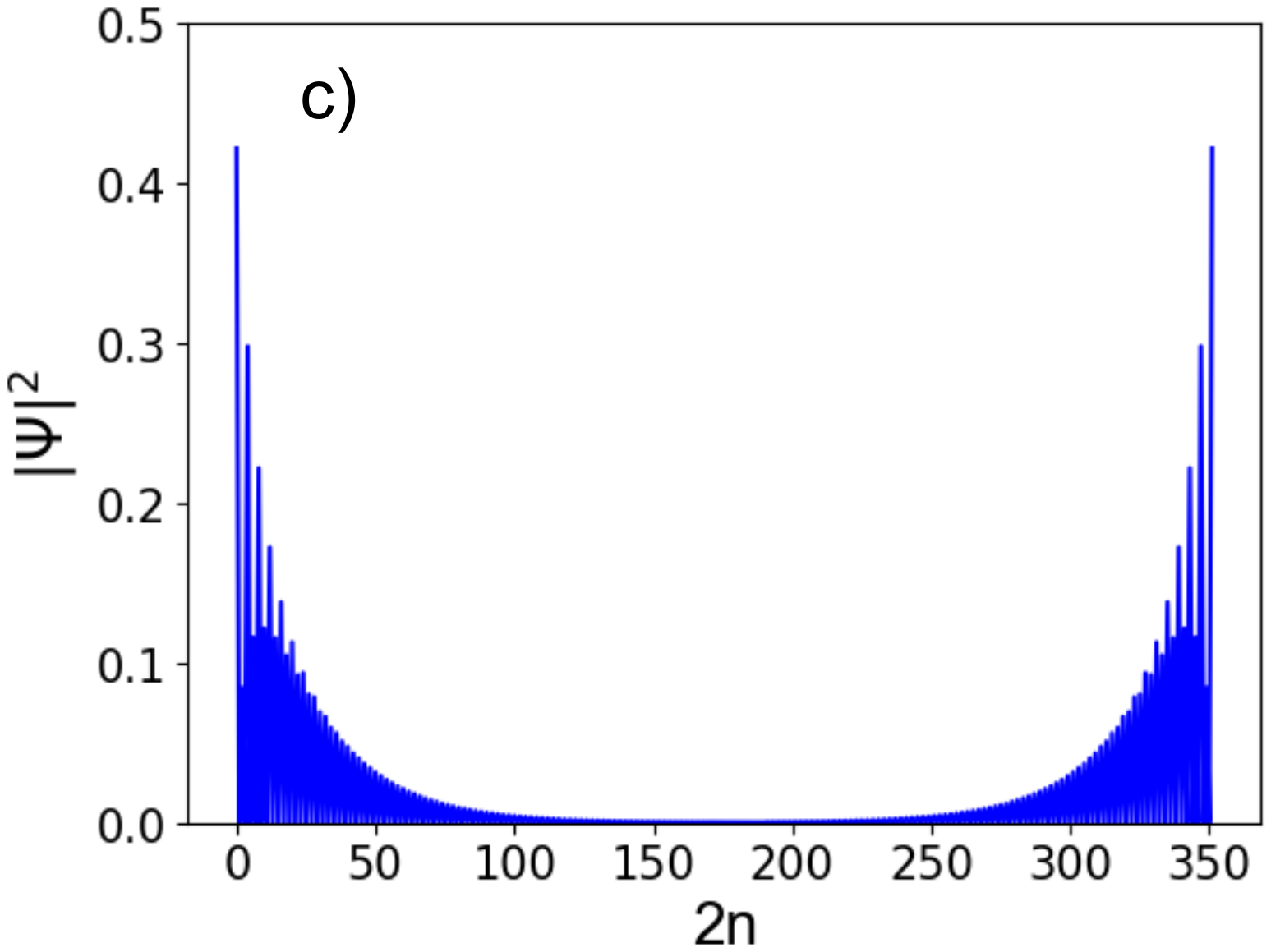}
\end{subfigure}
\caption{Amplitude square of the wave functions of the lower energy modes of a chain with 180 sites for fixed $\Delta/t=0.2$, with probability $p=1/2$ for a site  being occupied by a localized electron ($\langle n_f \rangle=1/2$) and  particular realizations of disorder. a) $J/t=0.1$. b) $J/t=0.3$. c) Amplitude of the wave-function of the ordered staggered configuration with $J/t=0.1$. Compare with a) and notice that in the random case the wave function is more localized.}
\label{fig8}
\end{figure}

For the homogeneous case, with a localized electron at each site of the chain there is no room for disorder. This is different for the occupation of the staggered configuration, $\langle n_f \rangle=1/2$, where the localized electrons may also  be randomly distributed in the chain. 
In the random case, with quenched disorder and $\langle n_f \rangle=1/2$, a given site has a probability $p=1/2$ of being empty or occupied by a local moment. It is interesting to explore the topological character of the random chains~\cite{desordem5,desordem3,desordem2}. Fig.~\ref{fig7} shows the eigenvalues for chains with 40 sites and 200 realizations of disorder for $p=1/2$.  We can observe the presence of zero energy modes  for the whole range of couplings, $J \le 1$, as shown in Fig.~\ref{fig7}.  Fig.~\ref{fig8}  shows the wave functions of these modes for some realizations of disorder and  different values of $J/t$ revealing that, in these cases, they are really edge  states. In  the ordered staggered configuration,  topological superconductivity occurs for $J/t<\Delta/t$.  In Fig.~\ref{fig8},  where  $\Delta/t=0.2$  is fixed, we notice   the existence of zero energy edge  states for $J/t=0.3>\Delta/t=0.2$. Then disorder extends the range of topological superconductivity to larger values of the coupling when compared to the ordered staggered configuration with the same occupation, $\langle n_f\rangle=1/2$.  This is a surprising effect that has been observed previously in rather similar systems~\cite{desordem5}. 

The problem of topological phases in random chains is complex and requires a closer examination. The existence of zero modes in chains with $p=1/2$ for  $(J/t) \le1$ does not necessarily imply the topological character of the chain in this region of the phase diagram. We find that in some configurations, with zero modes their wave-functions are not localized at the edges of the chain, but in the bulk. In order to establish the topological character of the random chains a more detailed analysis is necessary.  For this purpose we consider  a transfer matrix approach~\cite{gotta1,gotta2,derrida} that has been used in the study of localization in random chains with different types of disorder~\cite{derrida}. This allows to define a Lyapunov exponent that can distinguish between trivial and topological phases of the superconducting chain. 

The Heisenberg equation of motion for the Majorana operators $\alpha_A$  in the Hamiltonian, Eq.~\ref{HT} can be cast in the form of a recursion relation~\cite{gotta1,gotta2,gotta3},
$$
\begin{pmatrix}
 \alpha_{An+1}    \\
 \alpha_{An} 
 \end{pmatrix}
 = T_i
 \begin{pmatrix}
 \alpha_{An}    \\
 \alpha_{An-1} 
 \end{pmatrix},
$$
with a similar equation for the $\alpha_B$. The relevant, zero energy, site dependent  transfer matrix is given by,
\[T_i(E=0) =
\begin{pmatrix}
   \frac{-2J_i/t}{1+\Delta/t}      & \frac{-(1-\Delta/t)}{1+\Delta/t}  \\
  1     &  0 \\
\end{pmatrix}.
\]

Notice that we can write the interaction between localized and conduction electrons at a given  site $i$ of the chain, as $J_i=J x_i$ where $x_i=\pm 1$  according to the probability distribution, $P(x_i)=p \delta(x_i-1)+(1-p)\delta(x_i+1)$ for a given site.
\begin{figure}[tbh]
\centering
\includegraphics[width=1.\columnwidth]{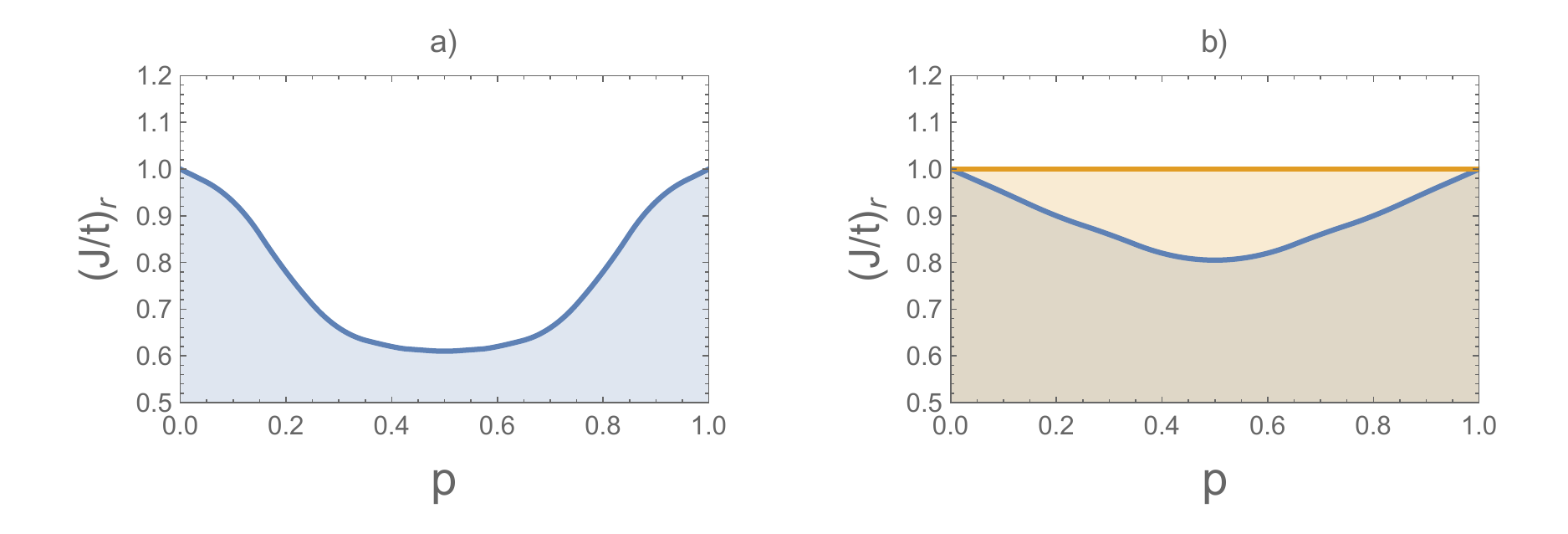}
\caption{(Color online) Phase diagram of chains with probability $p$ of  its  sites being occupied by a local moment. The lines in the figures represent the boundaries between trivial and topological phases and correspond to the points where the Lyapunov exponent $\gamma_r$,  changes sign. The  filled regions are topologically non-trivial, with  $\gamma_r<0$.   a) for $\Delta/t=0.2$. b) $\Delta/t=0.5$. In b) the horizontal orange line is the phase boundary for $\Delta/t=1$. }
\label{fig9}
\end{figure}
\begin{figure}[tbh]
\centering
\includegraphics[width=0.5\columnwidth]{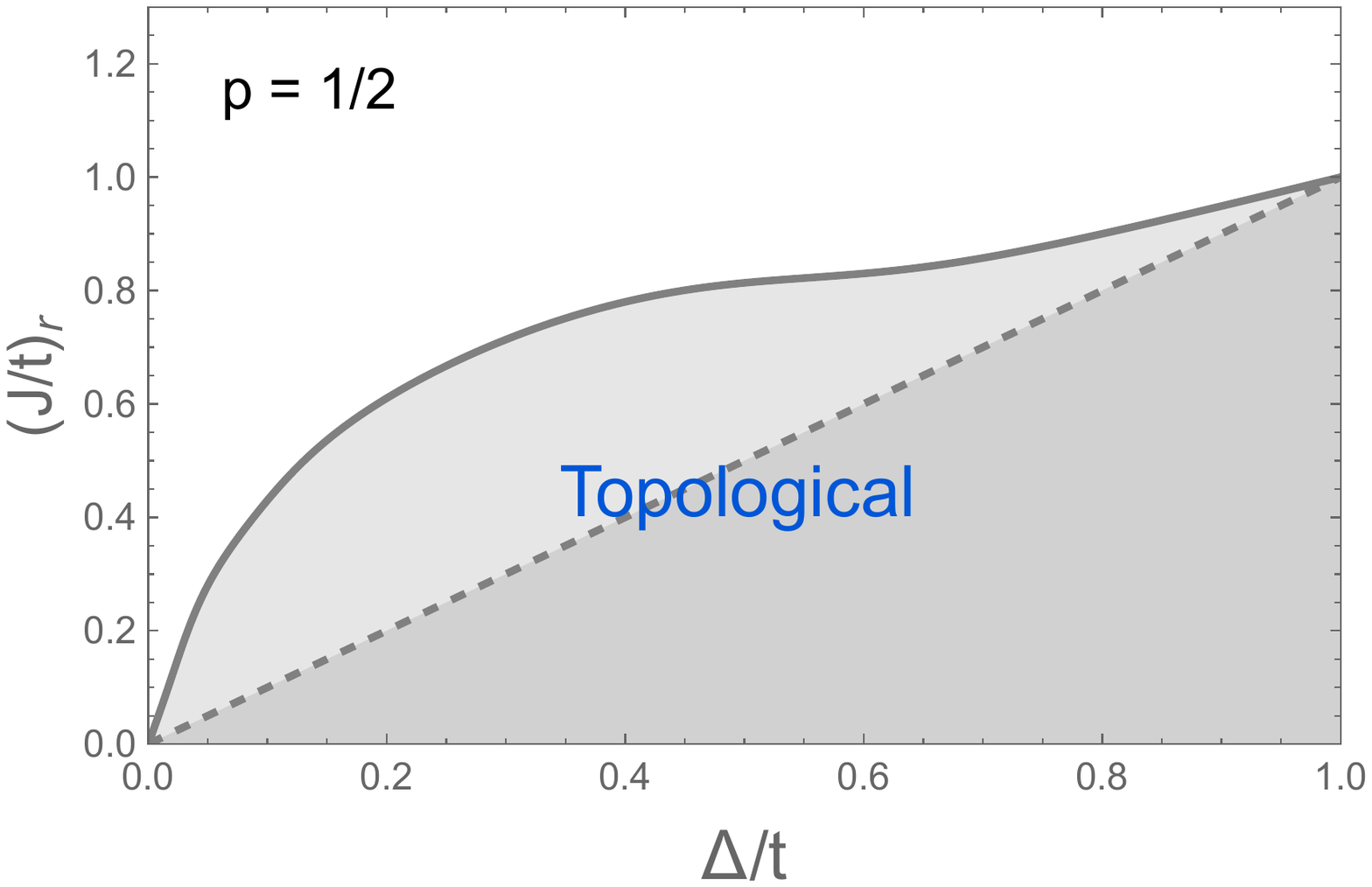}
\caption{(Color online) Phase diagram showing the critical coupling as a function of the gap amplitude for a fixed probability, $p=1/2$, of occupation of a site by a local moment. The straight,  dashed line is the phase boundary of the ordered staggered configuration. Disorder extends the topological region of the phase diagram. }
\label{fig10}
\end{figure}
The ratio $r_i=(\alpha_{A i+1}/\alpha_{Ai})$, as can be seen from the transfer matrix, obeys the following recursion relation~\cite{derrida},
\begin{equation}
\label{rr}
r_{i+1}=-\frac{2(J/t) x_i}{1+(\Delta/t)}-\frac{1-(\Delta/t)}{1+(\Delta/t)} \frac{1}{r_i}.
\end{equation}
This is a random map since the $x_i$ take random values in different sites.
We remark, the following relation involving the $r_i$,
\begin{equation}
\prod_{i=1}^N r_i = \frac{\alpha_{N+1}}{\alpha_N}\frac{\alpha_{N}}{\alpha_{N-1}} \frac{\alpha_{N-1}}{\alpha_{N-2}} \cdots \frac{\alpha_{1}}{\alpha_{0}}=\frac{\alpha_{N+1}}{\alpha_{0}},
\end{equation}
where we dropped the symbol $A$ from the operators $\alpha$. The quantity above describes the behavior of the amplitude of the wave functions of the edge modes as it penetrates in the chain. It can be used to define a Lyapunov exponent,
\begin{equation}
\label{lia2}
\gamma_r = \frac{1}{N} \sum_{i=1}^N \ln |r_i|.
\end{equation}
We have calculated numerically this Lyapunov exponent using Eq.~\ref{rr} to determine the phase diagrams of random chains with different values of $\Delta/t$, $J/t$ and $p$. The topological region of the phase diagram is characterized by  negative values of the Lyapunov exponents  and the phase boundary is where these change sign.
We carried out $15$  $\times$ $10^4$ iterations of Eq.~\ref{rr}  with $r_0=1$ to obtain  the phase diagrams shown in Figs.~\ref{fig9} and~\ref{fig10}.
Notice that  for $\Delta/t=1$, the Lyapunov exponent can be calculated exactly. Since the product of the $x_i$ can only assume the values, $\prod x_i=\pm 1$, we have $\gamma_r=\ln[|J/t|]$ independently of the probability $p$. This implies that the system is topological, whenever $J/t<1$, for any p.

We point out that  between the phase boundaries for $\Delta/t=1$ (yellow line) and $\Delta/t=1/2$ (blue line) in Fig.~\ref{fig9}b, we still observe zero modes in the spectra of eigenvalues, but the system is not topological in this region. For some of these modes we find that their wave functions are localized in the bulk of the chain.  
Finally,  Fig.~\ref{fig10} gives the phase diagram for a fixed value of $p=1/2$ showing the phase boundary for the disordered chain. In this figure the  straight line is the critical line for the ordered staggered configuration.  Notice that disorder increases the range of the topological phase in the phase diagram.

Finally,  we should mention that if we use an alternative approach,  and define a Lyapunov exponent from the product of the transfer matrices~\cite{gotta3},  we obtain similar results.

\section{Conclusions and perspectives}

In this work, we  studied the properties of the one-dimensional FK model with a $p$-wave pairing of the electrons in the conduction band.  We have shown that this many-body problem can be mapped in a non-interacting one, which allows it to be exactly soluble. The solution provides a clear picture of the effects of interactions and disorder in the topological properties of the superconducting chain. The conditions for solubility  are quite general, and permits to treat different numbers of localized electrons in the chain and arbitrary values of the FK interaction and superconducting pairing. In this paper we focused in a half-filled conduction band but the solution can be easily extended for general occupations.  

We find that for a chain with  a homogeneous configuration of localized electrons, i.e., with every site of the chain occupied by a local moment, the FK interaction leads eventually to a suppression of the superconducting topological phase, at large values of $J$ ($J_c=t$). In this system when turning on the pairing interaction, superconductivity arises from a metallic state of the FK model.  When the localized electrons assume a staggered  configuration, in which every other site in the chain is occupied, the critical value of the coupling to suppress the topological phase is much reduced, being given by the pairing energy of the conduction electrons ($J_c=\Delta$). In this case, superconductivity arises from an insulating state of the FK  model, and we can alternatively state  that a minimum critical value of the pairing amplitude, $\Delta_c=J$, is required to induce superconductivity in the insulator. 

The staggered configuration corresponds to $\langle n_f \rangle=1/2$. For this occupation  of local moments,  most of the configurations are disordered, with  the localized electrons  randomly occupying the sites in the chain. We have used transfer matrices and exact recursion relations to obtain the Lyapunov exponents and show 
that for $p=1/2$, randomness extends the range of stability of the topological superconducting phase with respect to the coupling to the localized electrons. While in the ordered staggered configuration,  the topological superconducting phase exists for $J/\Delta<1$, in the presence of disorder, it is stable up to larger values of $J$, as shown in  Fig.~\ref{fig10}.  

We have also studied  random configurations with an arbitrary probability $p$ for a site in the chain being occupied  by a localized electron.  The results obtained from the Lyapunov exponents yield the phase diagrams of the model and allows to put on firm grounds the role of interaction and  disorder in the topological properties of an archetypal system, namely the Kitaev chain. 

Finally, an interesting challenge is to include in this problem the hybridization between the localized and itinerant electrons, which gives rise to the quantum Falicov-Kimball model~\cite{brydon}.
We could also investigate the thermoelectric properties of these systems at the topological quantum phase transition of the homogeneous ($t=\Delta$) and staggered ($J=\Delta$), cases ~\cite{Andre}.

\begin{acknowledgments}
M A C acknowledges Brazilian National Council for Scientific and Technological Development (CNPq) Grant Number 305810/2020-0 and Foundation for Support of Research in the State of Rio de Janeiro (FAPERJ), Grant Number 201223/2021 for partial financial support.  M S F acknowledges financial support from the National
Council for Scientific and Technological Development (CNPq) Grant Number 311980/2021-0 and from the Foundation for Support of Research in the State of Rio de Janeiro (FAPERJ) process number 210 355/2018. A C P L acknowledges, Coordena\c{c}\~ao de Aperfei\c{c}oamento do Ensino Superior (CAPES) for financial support.
\end{acknowledgments}


\begin{thebibliography}{99}

\bibitem{alicea} Jason Alicea, New directions in the pursuit of Majorana fermions in solid state systems,  Rep. Prog. Phys. {\bf 75} 076501 (2012).
\bibitem{ando} Masatoshi Sato and Yoichi Ando,  Topological superconductors: a review, Rep. Prog. Phys. 80 076501 (2017).
\bibitem{review} M. M. Sharma, Prince Sharma, N. K. Karn and V. P. S. Awana, Comprehensive review on topological superconducting materials and interfaces, Supercond. Sci. Technol. 35 083003 (2022).
\bibitem{nature} Oliver Breunig and  Yoichi Ando, Opportunities in topological insulator devices, 
Nature Reviews Physics volume 4, pages 184 (2022).
\bibitem{superspin} Jacob Linder and Jason W. A. Robinson, Superconducting spintronics, Nature Physics, {\bf  11}, 307 (2015). 
\bibitem{kitaev} A. Y. Kitaev, Unpaired Majorana fermions in quantum wires, Physics-Uspekhi, 44, 131 (2001). 
\bibitem{Nayak1} C. Nayak, S. H. Simon, A. Stern, M. Freedman, and S. Das Sarma, Non-Abelian anyons and topological quantum computation, Reviews of Modern Physics {\bf 80}, 1083 (2008).
\bibitem{Nayak2} S. Das Sarma, M. Freedman, and C. Nayak, Majorana zero modes and topological quantum computation, npj Quantum Information {\bf 1}, 15001 (2015).
\bibitem{Refael} J. Alicea, Y. Oreg, G. Refael, F. von Oppen, and M. P. A. Fisher, Non-Abelian statistics and topological quantum information processing in 1D wire networks, Nature Physics {\bf 7}, 412 (2011).
\bibitem{Chou} P.-H. Chou, C.-H. Chen, S.-W. Liu, C.-H. Chung, and C.-Y. Mou, Geometry-induced topological superconductivity, Phys. Rev. B {\bf 103}, 014508 (2021).
\bibitem{Pan} H. Pan, S. Das Sarma, Majorana nanowires, Kitaev chains, and spin models, Phys. Rev. B {\bf 107}, 035440 (2023).
\bibitem{Protocol} M. Aghaee at al., InAs-Al Hybrid Devices Passing the Topological Gap Protocol (2022).
\bibitem{para2} Y. Oreg, G. Refael, and F. von Oppen, Helical Liquids and Majorana Bound States in Quantum Wires, Phys. Rev. Lett. {\bf 105}, 177002 (2010).
\bibitem{v8}  J. D. Sau, R. M. Lutchyn, S. Tewari, and S. Das Sarma, Generic New Platform for Topological Quantum Computation Using Semiconductor Heterostructures, Phys. Rev. Lett. 104, 040502 (2010).
\bibitem{v9} J. D. Sau, S. Tewari, R. M. Lutchyn, T. D. Stanescu, and S. Das Sarma, Non-Abelian quantum order in spin-orbit- coupled semiconductors: Search for topological Majorana particles in solid-state systems, Phys. Rev. B 82, 214509 (2010).
\bibitem{tewari2} S. Tewari, T. D. Stanescu, J. D. Sau, and S. Das Sarma, Topological minigap in quasi-one-dimensional spin-orbit-coupled semiconductor Majorana wires, Phys. Rev. B {\bf 86}, 024504 (2012).
\bibitem{para3} T. D. Stanescu and S Tewari, Majorana fermions in semiconductor nanowires: fundamentals, modeling, and experiment, J. Phys.: Condens. Matter {\bf 25} 233201 (2013).
\bibitem{bento} R. C. Bento Ribeiro, J. H. Correa,  L. S. Ricco,  A. C. Seridonio and M. S. Figueira, Spin-polarized Majorana zero modes in double zigzag honeycomb nanoribbons, Phys.Rev. {\bf B105}, 205115 (2022).
\bibitem{Robinson} J. Linder and J. W. A. Robinson, Superconducting Spintronics, Nat. Phys. {\bf 11}, 307 (2015).
\bibitem{Amitava} A. Bhattacharyya, D. Adroja, Y. Feng, D. Das, P. K. Biswas, T. Das, J. Zhao, $\mu$SR Study of Unconventional Pairing Symmetry in the Quasi-1D $\rm{Na_2Cr_3As_3}$ Superconductor. Magnetochemistry {\bf 9},70 (2023).
\bibitem{new2} E. Liebhaber, L. M. Rutten, G. Reecht,  et al. Quantum spins and hybridization in artificially-constructed chains of magnetic adatoms on a superconductor. Nat Commun {\bf 13}, 2160 (2022). 
\bibitem{tewari} E. Dumitrescu and S. Tewari, Topological properties of the time-reversal-symmetric Kitaev chain and applications to organic superconductors, Phys. Rev. {\bf B 88}, 220505(R) (2013).
\bibitem{desordem5} Niklas M. Gergs, Lars Fritz, and Dirk Schuricht, Topological order in the Kitaev/Majorana chain in the presence of disorder and interactions, Phys. Rev. B 93, 075129 (2016).
\bibitem{rachel} Stephan Rachel, Interacting topological insulators: a review,   Rep. Prog. Phys. 81 116501 (2018).
\bibitem{desordem3} G. Nunziante, A. Maiellaro, C. Guarcello, R. Citro,  Topological Phase Diagram of an Interacting Kitaev Chain: Mean Field versus DMRG Study. Condens. Matter  {\bf 9}, 20 (2024).
\bibitem{he} Yiting Deng, Yan He, Exact solutions of topological superconductor model with Hubbard interactions, Phys. Lett. {\bf A 397},127260 (2021).
\bibitem{desordem2} Max McGinley, Johannes Knolle, and Andreas Nunnenkamp, Robustness of Majorana edge modes and topological order: Exact results for the symmetric interacting Kitaev chain with disorder,
Phys. Rev. B 96, 241113(R) (2017).
\bibitem{desordem1}Haining Pan and S. Das Sarma, Disorder effects on Majorana zero modes: Kitaev chain versus semiconductor nanowire,
Phys. Rev. B 103, 224505 (2021).
\bibitem{desordem0} Simon Lieu, Derek K. K. Lee, and Johannes Knolle, Disorder protected and induced local zero-modes in longer-range Kitaev chains, Phys. Rev. B 98, 134507 (2018).
\bibitem{brydon} L.M. Falicov, J.C. Kimball, Simple model for semiconductor-metal transitions: SmB$_6$ and transition-metal oxides, Phys. Rev. Lett. 22, 997 (1969); P. M. R. Brydon and M. Gul\'acsi, Charge order in the Falicov-Kimball model, Phys. Rev. {\bf B73}, 235120 (2006);
T. Kennedy and E. H. Lieb, An itinerant electron model with crystalline or magnetic long range order, Physica  {\bf A138}, 320 (1986); U. Brandt and R. Schmidt, Exact results for the distribution
of the $f$-level ground state occupation in the spinless Falicov-Kimball model, Z. Phys. {\bf B63}, 45 (1986).
\bibitem{shen} Shen S.-Q., Topological insulators: Dirac equation in condensed matters, Vol. 174 (Springer) 2013.
\bibitem{novo} Christian Prosko, Shu-Ping Lee, and Joseph Maciejko, Simple $Z_2$ lattice gauge theories at finite fermion density, 
Phys. Rev. {\bf B96}, 205104 ( 2017).
\bibitem{molecule} C. Gruber,  D. Ueltschi, and J. Jsdrzejewski, Molecule Formation and the Farey Tree in the One-Dimensional Falicov-Kimball Model, Journal of Statistical Physics, Vol. 76, 125 (1994).
\bibitem{alase} Abhijeet Alase, {\it Boundary Physics and Bulk-Boundary Correspondence in Topological Phases of Matter}, Springer Theses, Springer Nature Switzerland AG, (2019).
\bibitem{gotta1} Wade DeGottardi, Diptiman Sen and Smitha Vishveshwara,Topological phases, Majorana modes and quench dynamics in a spin ladder system, New J. Phys., {\bf 13} 065028 (2011).
\bibitem{gotta2} Wade DeGottardi, Manisha Thakurathi, Smitha Vishveshwara, and Diptiman Sen, 
Majorana fermions in superconducting wires: Effects of long-range hopping, broken time-reversal
symmetry, and potential landscapes, Phys. Rev. {\bf B 88}, 165111 (2013) and references therein.
\bibitem{derrida} B. Derrida and E. Gardner, Lyapounov exponent of the one dimensional Anderson model : weak disorder expansions, J. Physique {bf 45}, 1283 (1984).
\bibitem{gotta3} Wade DeGottardi, Diptiman Sen, and Smitha Vishveshwara, Majorana Fermions in Superconducting 1D Systems Having Periodic, Quasiperiodic, and Disordered Potentials,
Phys. Rev. Lett. 110, 146404 (2013).
\bibitem{altland} see {\it Condensed Matter Field Theory}, Alexander Altland and Ben Simmons, Third Edition, Cambridge University Press, United Kingdom, 2023.
\bibitem{Andre} A. C. Lima,  R. C.  Bento Ribeiro, J. H. Correa,  Fernanda Deus, M. S. Figueira, and Mucio A. Continentino, Thermoelectric properties of topological chains coupled to a quantum dot, Scientific Reports 13, 1508 (2023).

\end{thebibliography}
\end{document}